\documentclass[%
 reprint,
 amsmath,amssymb,
 aps,
floatfix,
]{revtex4-2}
\usepackage{color} 
\usepackage{graphicx}
\usepackage{dcolumn}
\usepackage{bm}
\usepackage{physics, bm,bbm, dsfont}
\usepackage{mathrsfs}  
\usepackage{amsmath, amssymb, amscd, amsthm, amsfonts}
\usepackage{xparse}

\newcommand{\bxi}{\boldsymbol{\xi}}

\newcommand{\beeta}{\boldsymbol{\eta}}

\newcommand{\ee}{\text{e}}

\newcommand{\p}{\partial}
\newcommand{\bx}{\text{\bf x}}
\newcommand{\br}{\text{\bf r}}

\newcommand{\bo}{\text{\bf 0}}

\newcommand{\bsigma}{\boldsymbol{\sigma}}

\newcommand{\bnabla}{\boldsymbol{\nabla}}
\newcommand{\bmu}{\boldsymbol{\mu}}

\ExplSyntaxOn
\NewDocumentCommand{\eqrefs}{m}
 {
  \joansola_eqrefs:n { #1 }
 }

\seq_new:N \l__joansola_eqrefs_in_seq
\seq_new:N \l__joansola_eqrefs_out_seq

\cs_new_protected:Nn \joansola_eqrefs:n
 {
  \seq_set_split:Nnn \l__joansola_eqrefs_in_seq { , } { #1 }
  \seq_set_map:NNn \l__joansola_eqrefs_out_seq \l__joansola_eqrefs_in_seq
   {
    \exp_not:n { \ref{##1} }
   }
  \int_compare:nTF { \seq_count:N \l__joansola_eqrefs_out_seq < 2 }
   { Eq.\nobreakspace }
   { Eqs.\nobreakspace }
  \textup{(\seq_use:Nn \l__joansola_eqrefs_out_seq {,~})}
 }
\ExplSyntaxOff

\begin{document}

\preprint{APS/123-QED}

\title{Accelerating, to some extent, the $p$-spin dynamics}

\author{Federico Ghimenti}
 \email{federico.ghimenti@u-paris.fr}
\author{Frédéric van Wijland}%
\affiliation{Laboratoire Matière et Systèmes Complexes, Université de Paris and CNRS (UMR 7057), 10 rue Alice Domon et Léonie Duquet, 75013 Paris, France}%

\date{\today}

\begin{abstract}
We consider a detailed-balance violating dynamics whose stationary state is a prescribed Boltzmann distribution. Such dynamics have been shown to be faster than any equilibrium counterpart. We quantify the gain in convergence speed for a system whose energy landscape displays one, and then an infinite number of, energy barriers. In the latter case, we work with the mean-field disordered $p$-spin, and show that the convergence to equilibrium or to the nonergodic phase is accelerated, both during the $\beta$ and $\alpha$-relaxation stages. An interpretation in terms of trajectories in phase space and of an accidental fluctuation-dissipation theorem is provided.
\end{abstract}

\maketitle

\section{Markovian acceleration of the dynamics}
Designing algorithms able to locate the minima of a cost function in a large dimensional space is at the core of optimization problems lying well beyond the realm of physics. Yet the methods used, {\it e.g.} in computer science or applied mathematics, often borrow from physical approaches. Indeed the Boltzmann distribution $P_\text{\tiny B}\sim\ee^{-\beta H}$ for a system with energy $H$ concentrates, at low temperature $T=1/\beta$, in regions where $H$ is the smallest, and one of the important questions is thus whether one can efficiently sample the Boltzmann distribution in the low temperature regime, thereby locating the minima of $H$~\cite{kirkpatrick1983optimization}. This is achieved by endowing the degrees of freedom with dynamical evolution rules, often inspired by Langevin dynamics.

The question of how to fine-tune these dynamical evolution rules to accelerate the convergence to the steady state has grown into a field of its own, that is discussed at length in the physics literature as it touches to such diverse problems as the glass transition or finding the stable configuration of large proteins~\cite{mitsutake2001generalized}. It is of course abundantly discussed in the fields of optimization~\cite{welling2011bayesian} and machine learning~\cite{andrieu2003introduction} as well. For example, to accelerate the Stochastic Gradient Method, Nesterov~\cite{nesterov1983method} introduced a modified version of a momentum based ({\it a.k.a.} underdamped in the physics language) method. In the physics literature too, many cleverly accelerated dynamics have been introduced, such as the replica exchange~\cite{swendsen1986replica}, parallel tempering~\cite{marinari1992simulated} or the SWAP Monte Carlo method~\cite{grigera2001fast, berthier2016equilibrium, ninarello2017models}.

When designing alternative dynamics, it is also possible not to respect the detailed balance condition, now a widely used procedure in Monte Carlo schemes~\cite{suwa2010markov,dress1995cluster}, that leads to a faster convergence towards equilibrium~\cite{kapfer2017irreversible}. A particular subclass of algorithms is obtained by subjecting the system of interest to a divergenceless force such that the stationary state remains the Boltzmann distribution. This has been proposed by several authors over the last three decades~\cite{hwang1993accelerating,hwang2005accelerating, lelievre2013optimal,ichiki2013violation,ohzeki2015langevin,ohzeki2015mathematical,rey2015irreversible,rey2015variance,duncan2016variance}.
The terminology for a concrete implementation refers to a "lifting" procedure\cite{chen1999lifting,diaconis2000analysis,apers2017does} , which consists in introducing auxiliary degrees of freedom which drive the system away from equilibrium (but the statistics of which preserves the Boltzmann distribution for the degrees of freedom of interest). In the physics literature this was for instance demonstrated on the mean-field Ising model by \cite{turitsyn2011irreversible,vucelja2016lifting}.

In this work we shall build on the version proposed by Ichiki and Ohzeki~\cite{ohzeki2015langevin} (which we hereafter call the IO dynamics). They start from two independent and identical systems with a reversible dynamics whose equilibrium distribution $P_\text{\tiny B}$ is known. They then  add nonconservative solenoidal forces that couple the two systems but chosen such that $P_\text{\tiny B}$ remains the stationary distribution of the resulting dissipative dynamics for each individual system. The strength $\gamma$ of the coupling between the two copies of the system is a free parameter. As expected on general grounds,  the relaxation time to the stationary state is shorter in the presence of these peculiar forces. They have illustrated their findings on several simple examples. A key open question which we address in this work is to quantify the extent to which convergence to the Boltzmann distribution is accelerated in the presence of a single energy barrier, and in the presence of a large number of barriers in the energy landscape of the system of interest.

To answer these questions, we begin by asking how the IO dynamics behaves in the presence of a single energy barrier. This is a special case of the general study conducted by Bouchet and Reygner~\cite{bouchet2016generalisation} in which they determined a formal expression for the modified Kramers escape time. In addition to specializing the results of \cite{bouchet2016generalisation} we give a heuristic argument and we compare how the optimal trajectory deviates from the standard equilibrium counterpart. Then we implement the IO dynamics on a mean-field disordered $p$-spin system whose energy landscape is famous for harboring a very large number of metastable basins. We chose the disordered $p$-spin for the following reason: below a so-called dynamical temperature $T_d$, its equilibrium dynamics freezes and the resulting nonergodic behavior prevents the system from correctly sampling its target equilibrium Boltzmann distribution, so that it can be viewed as the most challenging system to test new ideas in optimization and sampling acceleration. We find that the dynamics is indeed accelerated, at various stages of the dynamics, though ergodicity breaking occurs at the same dynamical temperature. 

\section{The Ichiki-Ohzeki dynamics}

Let us begin by considering a particle with position $\br$ that evolves according to an overdamped Langevin dynamics in some energy landscape $V(\br)$ at temperature $T$. The equation of motion reads
\begin{equation}\label{eq:Langevindebase}
\dot{\br}=-\bnabla_\br V(\br)+\sqrt{2T}\beeta\end{equation}
where $\beeta$ is a Gaussian white noise with unit correlations $\langle\beeta(t)\otimes\beeta (t')\rangle=\mathds{1}\delta(t-t')$.  The companion Fokker-Planck equation 
\begin{equation}\label{eq:eqcur}
    \p_t P=-\bnabla_\br\cdot{\bf j},\,\,{\bf j}(\br,t)=-T\bnabla_\br P-\bnabla_\br VP 
\end{equation}
has the Boltzmann distribution $P_\text{\tiny B}(\br)\propto\ee^{-\beta V(\br)}$ as its stationary solution. Because of the underlying detailed balance condition, this stationary solution $P_\text{\tiny B}$ is actually an equilibrium one, as it makes the probability current $\bf j$ vanish, and no entropy is produced in the equilibrium steady-state. 

Consider now two identical particles with positions $\br^{(1)}$ and $\br^{(2)}$ evolving according to the coupled Langevin dynamics with independent white Gaussian noises $\beeta^{(1)}$ and $\beeta^{(2)}$,
\begin{align}\label{eq:Langevin-2}
\dot{\br}^{(1)}=-\bnabla_{\br^{(1)}} V(\br^{(1)})+\gamma\bnabla_{\br^{(2)}} V(\br^{(2)})+\sqrt{2T}\beeta^{(1)}\\
\dot{\br}^{(2)}=-\bnabla_{\br^{(2)}} V(\br^{(2)})-\gamma\bnabla_{\br^{(1)}} V(\br^{(1)})+\sqrt{2T}\beeta^{(2)}
\end{align}
The parameter $\gamma$ introduces a skew symmetric coupling between the two systems. Two independent overdamped Langevin dynamics for systems $1$ and $2$ are recovered when $\gamma = 0$. Note that when $\gamma \neq 0$ the zero temperature dynamics of a single system is not a gradient descent down its potential landscape. Yet a gradient descent dynamics for the composite system is recovered, since by denoting $U(\br^{(1)},\br^{(2)}) \equiv V(\br^{(1)}) + V(\br^{(2)})$ we have 
\begin{equation}
    \frac{\dd}{\dd t}U(\br^{(1)},\br^{(2)}) = -\sum_{a=1,2} \left(\bnabla_{\br^{(a)}} V(\br^{(a)})\right)^2 < 0
\end{equation}
 Therefore, should one of the subsystems get stuck in some local minimum, then the other one is able to temporarily increase its energy. The latter system then acts as a source of fluctuations for the stuck system. At nonzero temperature, these fluctuations add to the thermal ones, allowing for a faster exploration of the configuration space. 
 
Let's verify that the nonequilibrium stationary state measure is the Boltzmann measure. The probability current in the Fokker-Planck equation for $P(\br^{(1)},\br^{(2)},t)$ follows from Eq.~\eqref{eq:Langevin-2}. With respect to the expression of Eq.~\eqref{eq:eqcur} picks up an additional contribution for each of the two subspaces, namely 
\begin{equation}\begin{split}
\delta{\bf j}^{(1)}= \gamma \bnabla_{\br^{(1)}} V(\br^{(2)})P(\br^{(1)},\br^{(2)},t)\\
\delta{\bf j}^{(2)}=-\gamma\bnabla_{\br^{(2)}} V(\br^{(1)})P(\br^{(1)},\br^{(2)},t)
\end{split}\end{equation}
This additional current arises from the presence of the nonconservative forces in Eq.~\eqref{eq:Langevin-2}, and it is divergence-free when $P(\br^{(1)},\br^{(2)},t)=P_\text{\tiny B}(\br_1)P_\text{\tiny B}(\br_2)$. We therefore know that this factored Boltzmann distribution is the stationary state of the coupled process. However the dynamical rules in Eq.~\eqref{eq:Langevin-2} are accompanied by a positive entropy production rate $\dot{\Sigma}$ given by
 \begin{equation}\label{eq:entropyprod}
\dot{\Sigma}= 2\beta\gamma^2\langle(\bnabla V)^2\rangle_\text{\tiny B}=2\gamma^2\langle\Delta V\rangle_\text{\tiny B}>0
 \end{equation}
 where $\langle\ldots\rangle_\text{\tiny B}$ refers to an average with respect to the Boltzmann distribution, thus confirming the nonequilibrium nature of the dynamics, and that $P_\text{\tiny B}$ is a nonequilibrium steady-state distribution. This calculation tells us about the typical time scale $1/\dot{\Sigma}$ it takes for the dynamics to display irreversible features.
 
 We now give a physicist's argument explaining why the IO dynamics is faster than the standard Langevin dynamics of Eq.~\eqref{eq:Langevindebase}.

\subsection{A simple argument for the decrease  of the relaxation time}

A mathematical proof accounting for the decrease of the relaxation time for arbitrary $\gamma$ was given in \cite{hwang1993accelerating, hwang2005accelerating} and it was rediscovered in the case of discrete state Markov processes in \cite{ichiki2013violation}. However, for small $\gamma$, there is a simple physicist's argument to account for the increase in the relaxation rate towards the steady state of the Fokker-Planck operator. To see this we use the Darboux~\cite{darboux1946leccons} and we rewrite the Fokker-Planck operator $\mathbb{W}$ associated to Eq. \eqref{eq:Langevin-2} in a rotated basis, $\tilde{\mathbb{W}}=P_\text{\tiny B}^{1/2}\mathbb{W}P_\text{\tiny B}^{-1/2}$, which reads
\begin{equation}\label{eq:FPopDarboux}
    \tilde{\mathbb{W}}=\tilde{\mathbb{W}}_0^{(1)}+\tilde{\mathbb{W}}_0^{(2)}+\delta\tilde{\mathbb{W}} 
\end{equation}
where 
\begin{equation}
    \begin{split}
        \tilde{\mathbb{W}}_0^{(a)} &\equiv T\bnabla_{\br^{(a)}}^2 \\
        &-\frac{1}{4T}\left[(\bnabla_{\br^{(a)}}V(\br^{(a)}))^2-2T\bnabla^2_{\br^{(a)}}V(\br^{(a)})\right]
    \end{split}    
\end{equation}
is the symmetrized Fokker-Planck operator for particle $a=1,2$ in the absence of the coupling parameter $\gamma$ and
\begin{equation}
    \delta\tilde{\mathbb{W}} \equiv \gamma\left[ \left(\bnabla_{\br_1}V(\br_1)\right)\cdot\bnabla_{\br_2}-\left( \bnabla_{\br_2}V(\br_2)\right)\cdot\bnabla_{\br_1}\right]
\end{equation}
As expected, for $\gamma=0$ this is a Hermitian operator (which reflects the detailed balance condition), but for $\gamma \neq 0$ a skew-Hermitian perturbation is introduced in Eq.~\eqref{eq:FPopDarboux}. We denote by $\ket{\Psi_n}$ the eigenstates of the unperturbed Hermitian operator $\tilde{\mathbb{W}}_0^{(1)} + \tilde{\mathbb{W}}_0^{(2)}$, and with $E_n$ the associated eigenvalue. We are interested in the change $\delta E$ to the spectral gap $\Delta E = E_1 - E_0$ of the operator $-\tilde{\mathbb{W}}$ between the ground state $\ket{\Psi_0}$ and the first excited state $\ket{\Psi_1}$. We first observe that the ground state and its (zero) energy are left unchanged by the perturbation. Therefore, to second order in perturbation theory, we have
\begin{equation}\label{eq:order2qpt}
    \begin{split}
        \Delta E &=- \gamma \mel{\Psi_1}{\delta\tilde{\mathbb{W}}}{\Psi_1} \\
                &- \gamma^2\sum_{n>1}\frac{\abs{\mel{\Psi_1}{\delta\tilde{\mathbb{W}}}{\psi_n}}^2}{E_1 - E_n}
    \end{split}
\end{equation}
Due to the skew-Hermiticity of the perturbation the first term in Eq.~\eqref{eq:order2qpt} is imaginary, and the second term is real and positive. To second order in $\gamma$, the eigenvalues of $\tilde{\mathbb{W}}$ are shifted in the opposite direction to what would occur in the case of a Hermitian perturbation. Only the real part of the spectral gap is relevant in the computation of the relaxation times, and since it is increased by the perturbation, the relaxation time toward the steady state distribution is reduced.

This increase in the relaxation rate has been explored for simple systems in \cite{ohzeki2015langevin,ohzeki2015mathematical,ichiki2021stochastic}, but for systems whose dynamics is controlled by rare---yet important---events, the efficiency of the IO approach deserves deeper investigation. As a first step, we revisit Kramers' escape problem with the IO dynamics Eq.~\eqref{eq:Langevin-2} to understand qualitatively how it copes with a single energy barrier.\\

\section{The barrier crossing problem}\label{sec:kramers}

Consider a one-dimensional potential landscape $V(x)$ with a stable equilibrium position at $x_m$  and ask about the time $\tau$ it takes for particle $1$ to hop over a potential barrier $\Delta V$, passing through the local maximum at $x_M$, starting from the bottom of the well. As we know from Kramers' analysis~\cite{kramers1940brownian}, for equilibrium dynamics, $\tau$ is given by the Arrhenius formula $\tau\simeq\ee^{\beta \Delta V}$. Consider now a nonzero $\gamma$ for two coupled particles. Both particles start from the bottom of the potential well. We ask how long it takes particle 1 to hop over the potential barrier irrespective of the motion of particle 2. We begin our presentation with a heuristic analysis before connecting to a recent work by Bouchet and Reygner~\cite{bouchet2016generalisation}.

\subsection{Heuristic argument}\label{sec:heuristic}

Before proceeding with a detailed analysis, there is a simple way to understand how a nonzero $\gamma$ speeds barrier crossing up. We are interested in the rare event in which particle $1$ will hop over an energy barrier while, at low temperature, the typical motion of particle 2 remains confined within the bottom of the well. The position of particle 2 then evolves according to
\begin{equation}\begin{split}
\dot{x}^{(2)}(t)=&-V'(x^{(2)}(t))-\gamma V'(x^{(1)}(t))+\sqrt{2T}\eta^{(2)}(t)\\
\simeq&-k (x^{(2)}(t)-x_m)-\gamma V'(x^{(1)}(t))+\sqrt{2T}\eta^{(2)}(t)
\end{split}
\end{equation}
where  $k_m \equiv V''(x_m)>0$ is the stiffness of the well. We can integrate this equation and substitute $x^{(2)}$ into the evolution equation for $x^{(1)}$:
\begin{equation}\label{AOUP}
    \begin{split}
        \dot{x}^{(1)}(t)=&-\int_0^t\dd s M_R(t-s)V'(x^{(1)}(s))+\sqrt{2T}\eta(t)\\
    \end{split}
\end{equation}
where $\eta(t)$ is now a colored Gaussian noise with correlations $\langle\eta(t)\eta(t')\rangle=M_C(t-t')$. The mobility kernel $M_R(t)=\delta(t)+\gamma^2 k_m\ee^{-k_m t}\theta(t)$ and the correlation kernel $M_C(t)=\frac 12 M_R(t)+\frac 12 M_R(-t)$ are related by Kubo's fluctuation theorem for generalized Langevin equations. While the two-dimensional process $(x^{(1)},x^{(2)})$ is a nonequilibrium one, the effective process $x^{(1)}$, forgetting the information on $x^{(2)}$, is a truly equilibrium process (at least when $x^{(2)}$ evolves in a quadratic well). In the large $k_m$ limit, $M_R(t)\simeq (1+\gamma^2)\delta(t)$ and the mobility of the particle picks up a $\gamma^2$ contribution so that the Kramers' escape time is reduced accordingly: $\tau\simeq \ee^{\beta\Delta V}/(1+\gamma^2)$. 

This heuristic approach suggests that the IO dynamics contributes to changes in the barrier crossing time by rescaling the prefactor of the Arrhenius formula. We will confirm this observation in the general case by looking at the most probable path taken by the system during the barrier crossing process, also known as the instanton trajectory.

To do so, it is convenient to cast the equations of motion for the two coupled one-dimensional processes in a matrix form
\begin{equation}
     \mathbf{\dot x} = -\left( \mathds{1} + \bm{\Gamma} \right) \cdot \bnabla U(\mathbf{x}) + \sqrt{2T}\bm{\eta}(t)
\end{equation}
 with $\mathbf{x} \equiv \begin{bmatrix} x^{(1)} & x^{(2)} \end{bmatrix}^T$, $U(\mathbf{x}) \equiv V(x^{(1)}) + V(x^{(2)})$ and 
\begin{equation}
    \bm{\Gamma} \equiv \begin{bmatrix}
                            0 & -\gamma \\
                            \gamma & 0
                        \end{bmatrix}
\end{equation}  
This two-dimensional system presents a local minimum at $\mathbf{x}_m \equiv \begin{bmatrix} x_m & x_m \end{bmatrix}^T$ and a saddle point at $\mathbf{x}_s \equiv \begin{bmatrix} x_M & x_m\end{bmatrix}^T$. The process in which particle $1$ overcomes the energetic barrier while particle $2$ remains confined in the well corresponds, for the composite system, to a trajectory from $\mathbf{x}_m$ to $\mathbf{x}_s$ and from $\mathbf{x}_s$ to the minimum on the other side of the saddle. We now proceed with a careful analysis of the instanton trajectory.

\subsection{Instanton trajectory: no change in the exponential factor}

\begin{figure}
    \centering
    \includegraphics[width = .4\textwidth]{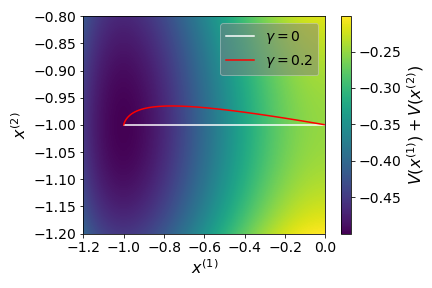}
    \caption{Instanton trajectory from Eq. \eqref{eq:instanton} for $\gamma = 0$ and its perturbative expansion up to second order in $\gamma$ for the double well potential $V(x) = \frac{1}{4}x^4 - \frac{1}{2} x^2$.}
    \label{fig:instanton}
\end{figure}

The Kramers escape rate $\tau^{-1}$ can be expressed within a path-integral formulation,
\begin{equation}
        \tau^{-1}=\int {\mathscr D}\mathbf{\hat{x}}{\mathscr D}\bx\ee^{-\frac{1}{T}S\left[\mathbf{x}, \mathbf{\hat{x}}\right]}
\end{equation}
where the path-integral runs over paths connecting $\mathbf{x}_m $ to $\mathbf{x}_s$, and where the Janssen-De Dominicis action functional reads \begin{equation}\label{eq:actionJdD}
    S\left[\mathbf{x}, \mathbf{\hat{x}} \right] = \int_{-\infty}^{+\infty} \dd t\left[ \mathbf{\dot x}\cdot\mathbf{\hat{x}} - H\left(\mathbf{x}, \mathbf{\hat{x}} \right)\right]
\end{equation}
The response field $\mathbf{\hat{x}}$ expresses the effect of the noise on the dynamics of $\mathbf{x}$. In the small noise $T \rightarrow 0$ limit, the path that dominates the integral minimizes $S$, and the resulting saddle-point trajectories follow a Hamiltonian-like dynamics (with respect to $H$ in Eq.~\eqref{eq:actionJdD}) in a space of extended variables where $\mathbf{\hat{x}}$ stands for a momentum. The escape time $\tau$ then satisfies the logarithmic equivalence 
\begin{equation}
    \begin{cases}
        \tau \asymp \ee^{\frac{\phi}{T}} \\
        \phi \equiv \min_{\mathbf{x}(t) | \mathbf{x}(-\infty) = \mathbf{x}_m \text{, } \mathbf{x}(+\infty) = \mathbf{x}_s} S\left[\mathbf{x}, \mathbf{\hat{x}}\right]
    \end{cases}
\end{equation}
The explicit expression for the Hamiltonian $H(\mathbf{x},\mathbf{\hat{x}})$ is
\begin{equation}
    H(\mathbf{x},\mathbf{\hat{x}}) \equiv - \mathbf{\hat{x}} \cdot \left( \mathds{1} + \bm{\Gamma} \right)\cdot \bnabla U(\mathbf{x}) + \mathbf{\hat{x}}^2 
\end{equation}
The minimum of $S$ for the trajectories of interest is found by looking for the solutions of the Hamiltonian system induced by $H$ along the $H=0$ energy manifold, so that we have the differential equations
\begin{equation}\label{eq:saddlepoint}
    \begin{cases}
        \mathbf{\dot x} &= \frac{\p H}{\p\mathbf{\hat{x}}}=-\left( \mathds{1} + \bm{\Gamma} \right)\cdot \bnabla U(\mathbf{x})  + 2\mathbf{\hat{x}} \\
        \mathbf{\dot{\hat{x}}} &=-\frac{\p H}{\p\mathbf{x}}= \left(\mathds{1} + \mathbf{\Gamma}\right) \cdot \text{Hess } U(\mathbf{x}) \cdot \mathbf{\hat{x}}
    \end{cases}
\end{equation}
together with the condition $H(\mathbf{x},\mathbf{\hat{x}}) = 0$. Here $\text{Hess } U(\mathbf{x})_{ab} \equiv V''(x^{(a)})\delta_{ab}$ is the Hessian matrix of $U$. These coupled differential equations are equivalent to
\begin{equation}\label{eq:instanton}
    \begin{cases}
        \mathbf{\dot x} &= \left(\mathds{1} - \mathbf{\Gamma}\right) \cdot \mathbf{\bnabla}U(\mathbf{x}) \\
        \mathbf{\hat x} &= \mathbf{\bnabla}U(\mathbf{x})
    \end{cases}
\end{equation}
which, in principle, can be solved explicitly if a potential $V$ is specified. We solve the instanton trajectory $\dot{\mathbf{x}}$ within perturbation theory up to second order in $\gamma$ as a functional of the external potential $V$, assuming the $\gamma=0$ trajectory is fully known. A linear analysis of the trajectory close to the saddle point $\mathbf{x}_s$ shows (see App.~\ref{app:instantonlinearised}) that the perturbation expansion must have the form
\begin{equation} \label{eq:perturbationtheory}
    \begin{aligned}
        x^{(1)} &= x^{(1)}_0 + \gamma^2 x^{(1)}_2 + \mathcal{O}(\gamma^4) \\
        x^{(2)} &= x^{(2)}_0 + \gamma x^{(2)}_1 + \mathcal{O}(\gamma^3)
    \end{aligned}
\end{equation}
with $\dot{x}^{(1)}_0 = V'\left(x^{(1)}_0(t)\right)$ and $x^{(2)}_0 = 0$. By taking into account the boundary condition for the perturbation $x(+\infty)^{(a)}_i = x(-\infty)^{(a)}_i = 0$ for $a = 1,2$, $i > 0$ we obtain
\begin{align}
    x^{(2)}_1 &= -\ee^{k_m t}\int_{t}^{+\infty} \dd\tau\; \ee^{-k_m \tau}V'\left(x^{(1)}_0(\tau)\right) \\
    x^{(1)}_2 &= k_m \ee^{G(t)}\left[\int_{t}^{+\infty}\dd\tau\; \ee^{-G(\tau)}x^{(2)}_1(\tau) + C \right]
\end{align}
where $G(t) \equiv \int \dd\tau V''(x^{(1)}_0(\tau))$ and $C$ is a constant determined by matching the asymptotic expression for $x^{(1)}_2(t)$ for $t \rightarrow +\infty$ to the linearized expansion at $\mathbf{x}_s$. We illustrate the role played by $\gamma$ in deforming the equilibrium trajectory in Fig.~\ref{fig:instanton}. As done in previous studies of the escape time out of equilibrium~\cite{bray1989instanton}, we have chosen as an explicit potential $V$ the well-known double well $V(x) \equiv \frac{1}{4}x^4 - \frac{1}{2}x^2$. We see that indeed system 2 finds it optimal to rise in its own energy landscape to favor the crossing of system 1 over the barrier. 
Interestingly, the action for the instanton trajectory is unchanged with respect to the equilibrium dynamics, namely
\begin{equation}
     S\left[\mathbf{x}, \mathbf{\hat{x}} \right] = \int_{-\infty}^{+\infty} \dd t \mathbf{\dot x} \cdot \mathbf{\bnabla}U(\mathbf{x}) = \Delta U
\end{equation}
where $\Delta U = V(x_M) - V(x_m)$. That this result is identical to the one obtained with the equilibrium dynamics at $\gamma = 0$ supports the the heuristic analysis of Sec. \ref{sec:heuristic}. While the instanton trajectory is clearly altered, the Arrhenius factor is left unchanged by the IO dynamics. In the next section, we will investigate the role of $\gamma$ on the prefactor to the Arrhenius exponential.

\subsection{Full expression of $\tau$}

Bouchet and Reygner~\cite{bouchet2016generalisation} have derived an Eyring-Kramers formula for nonequilibrium dynamics which do not necessarily sample the Boltzmann distribution. Here we specialize their general result to the IO dynamics, which allows us to arrive at a concrete $\gamma$-dependence of the escape time. When applied to the IO dynamics the time $\tau$ reads
\begin{equation}\label{eq:bouchetreygner}
    \tau = \frac{2\pi}{\lambda_{+}}\sqrt{\frac{k_M}{k_m}} \exp\left(\frac{\Delta V}{T}\right)
\end{equation}
where $k_M \equiv -V''({x}_M) > 0$, $k_m \equiv V''(x_m) > 0$, and $\lambda_+$ is the modulus of the eigenvalue related to the unstable direction of the dynamics linearized around the saddle point $\mathbf{x}_s$:
\begin{equation}\label{eq:lambdaplus}
    \lambda_+ = \frac{1}{2}\left[ k_M - k_m + \sqrt{\left(k_M + k_m\right)^2 + 4\gamma^2k_M k_m}\right] 
\end{equation}
For $\gamma = 0$ Eq. \eqref{eq:bouchetreygner} yields the well-known Kramers' time
\begin{equation}\label{eq:kramerIO}
    \tau = \frac{2\pi}{\sqrt{k_M k_m}} \exp\left(\frac{\Delta V}{T}\right)
\end{equation}
while for $\gamma \neq 0$ we have $\lambda_+ > k_M$ and the time required for particle $1$ to hop over the energetic barrier is reduced.  

We realize {\it a posteriori} that the heuristic approach of Sec. \ref{sec:heuristic} was physically sound. It can indeed be recovered by letting $k_m \rightarrow +\infty$ in Eq.~\eqref{eq:lambdaplus}, yielding $\lambda_+ \approx k_M (1 + \gamma^2)$. This limiting case is actually the smallest achievable value of $\tau$ at given $\gamma$, $k_M$ and $\Delta V$, since $\partial_{k_m} \lambda_+$ is always positive.\\ 

 The results of this section precisely quantify the gain in speed that $\gamma$ leads to in the crossing of a single steep energy barrier. A natural question that arises next is whether any similar gain --to be quantified as a function of $\gamma$--- can be harvested in the presence of  a  large number of steep energy barriers. We do not expect the IO dynamics to overcome the ergodicity breaking that occurs in some systems where the energy barriers become infinitely high, as is for instance the case in some mean-field disordered system in the large system size limit. In what follows, our goal is to put the IO dynamics to a severe test. We shall now investigate how it performs on a specific $N$-body system known for displaying strong ergodicity breaking behavior at low temperatures, the disordered $p$-spin model for $p\geq 3$.
 
\section{$p$-spin with Ichiki-Ohzeki dynamics}
The spherical $p$-spin is a disordered fully connected model of $N$ interacting spins $\sigma_1,\ldots,\sigma_N$ fulfilling the constraint $\sum_{i=1}^N\sigma_i^2=N$. The energy of a configuration is
\begin{equation}\label{eq:Hpspin}
{\mathscr H}[\{\sigma_i\}_{i=1,\ldots,N}]=-\sum_{i_1<\ldots<i_p}J_{i_1\ldots i_p}\sigma_{i_1}\ldots\sigma_{i_p}
\end{equation}
where the $J_{i_1\ldots i_p}$ are independent quenched random couplings with variance $\frac{2J^2}{p!N^{p-1}}$. To cut a long story short, we refer the reader to the huge body of literature~\cite{crisanti1992spherical, crisanti1993sphericalp, barrat1996dynamics,barrat1997p, castellani2005spin, de2006random} devoted to the $p$-spin to understand the role it has played in shaping up our understanding of disordered and glassy systems (including active ones~\cite{berthier2013non}). A summary of the properties relevant to the present discussion is that this model undergoes a transition from a paramagnetic state at high temperatures to a spin-glass phase at low temperatures $T<T_c$~\cite{crisanti1992spherical}. From a static point of view, its peculiarity is that there exists an intermediate temperature $T_d>T_c$ such that in the temperature range $T_c<T<T_d$ an exponentially large number of metastable states appears, without however affecting the paramagnetic nature of the system. From a dynamical standpoint, while the system relaxes above $T_d$, it fails to explore this complex energy landcsape as soon as $T<T_d$~\cite{crisanti1993sphericalp}. When endowed with purely relaxational equilibrium dynamics, the spin-spin correlation function $C(t,t')=\frac{1}{N}\sum_i\langle\sigma_i(t)\sigma_i(t')\rangle$, in the regime with $t'$ large and $t-t'$ finite with time-translation invariance, fails to relax to $0$ for $T<T_d$, reaching instead a temperature-dependent plateau value $q$. The $p$-spin is thus the ideal test-bench to quantify the acceleration of the IO dynamics. 
\subsection{Mean field IO dynamics}
In order to implement the IO dynamics we consider two coupled $p$-spin systems (with the same realization of the disorder), with spins $\sigma_i^{(1)}$ and $\sigma_i^{(2)}$ evolving according to
\begin{align}\label{eq:Langevinpspin}
    \begin{split}
        \p_t\sigma_i^{(1)}&=-\frac{\p {\mathscr H}}{\p\sigma_i^{(1)}}+\gamma\frac{\p {\mathscr H}}{\p\sigma_i^{(2)}} \\
        &-\mu^{(1)}(t)\sigma_i^{(1)} + \gamma \mu^{(2)}(t)\sigma_i^{(2)}+\sqrt{2T}\eta_i^{(1)}
    \end{split}\\
    \begin{split}
        \p_t\sigma_i^{(2)}&=-\frac{\p {\mathscr H}}{\p\sigma_i^{(2)}}-\gamma\frac{\p {\mathscr H}}{\p\sigma_i^{(1)}}\\ 
        &-\mu_2(t)\sigma_i^{(2)} - \gamma\mu^{(1)}(t) \sigma_i^{(1)} + \sqrt{2T}\eta_i^{(2)}
    \end{split}
\end{align}
where the functions $\mu^{(a)}(t)$ are adjusted so that the spherical constraint $\sum_{i=1}^N\left\langle \left(\sigma_i^{(a)}(t)\right)^2 \right\rangle=N$ is fulfilled throughout the dynamical evolution. There is a large body of literature on the effect of a nonequilibrium drive on the $p$-spin (or on other complex systems with a similar glassy behavior at low temperatures). In these works  it has consistently been found that in the presence of a nonequilibrium drive (be it in a randomized non-relaxational dynamics~\cite{cugliandolo1997glassy}, or in a shear~\cite{berthier2000two} or with an active force~\cite{berthier2013non}) a lower temperature than the equilibrium one is required in order to witness the onset of ergodicity breaking. However, in those cases the nonequilibrium stationary-state that is being sampled is not the Boltzmann distribution, and it is in general unknown. We stress that, unlike in the latter series of works, the IO dynamics does correctly sample, in its stationary state, the Boltzmann distribution $P_\text{\tiny B}\sim\ee^{-\beta {\mathscr H}}$ for ${\mathscr H}={\mathscr H}[\{\sigma_i^{(1)}\}]+{\mathscr H}[\{\sigma_i^{(2)}\}]$ (with ${\mathscr H}$ given in Eq.~\eqref{eq:Hpspin}), despite being a nonequilibrium dynamics. From an analytical standpoint, a remarkable feature of the standard dynamics of the $p$-spin, which is preserved by Eq.~\eqref{eq:Langevinpspin} is that the mean-field nature of the problem allows us to write the stochastic evolution of a single spin in terms of unknown functions (the spin-spin correlation, the response function that enters the memory kernel, and noise correlations) that are determined self-consistently. Such a procedure can be applied to our case. Note however that in the context of the IO dynamics, the spin-spin correlation becomes a collection of four spin-spin correlations $C_{ab}(t,t')=\langle\sigma^{(a)}(t)\sigma^{(b)}(t')\rangle$, and the same holds for the response function $R_{ab}(t,t')$ and the integrated response $F_{ab}(t,t') \equiv -\int_{t'}^t \dd\tau R_{ab}(t,\tau)$ . In practice one is able to write an effective Langevin equation for the dynamics of a single spin $\sigma^{(1)}$ in system $1$ and its counterpart $\sigma^{(2)}$ in system $2$, from which self-consistent equations of motion for the correlation and the integrated response can be derived. Since our interest goes to the dynamics in the steady state we assume time translational invariance to hold, $C_{ab}(t,t') = C_{ab}(t-t')$ and $F_{ab}(t,t') = F_{ab}(t-t')$, and by applying standard path integral techniques we arrive at the following Langevin equation for $\bsigma(t)\equiv \begin{bmatrix} \sigma^{(1)} & \sigma^{(2)}) \end{bmatrix}^T$ and the dynamical equations for the correlation matrix $\mathbf{C}$ and the integrated response matrix $\mathbf{F}$:
\begin{widetext}
\begin{align}
  \p_t {\bsigma}(t) &= -\left(\mathds{1} + \bm{\Gamma}\right) \cdot {\bmu}(t) \cdot {\bsigma}(t) + \int_0^t \dd\tau \mathbf{M_F}(t-\tau) \cdot {\bsigma}(\tau)
    + \frac{1}{T}\left(\mathds{1} + \bm{\Gamma} \right) \cdot \mathbf{C}_0(p,t) \cdot {\bsigma}(0) + {\bxi}(t) \label{eq:effestocheq}\\
        \partial_t\mathbf{F}(t) &= - \mathds{1} - \left(\mathds{1} + \bm{\Gamma}\right) \cdot \bm{\mu}(t) \cdot \mathbf{F}(t) + \int_{t'}^t d\tau \mathbf{M_F}(t-\tau) \cdot \mathbf{F}(\tau) \label{eq:F}\\
        \partial_t \mathbf{C}(t) &= -\left(\mathds{1} + \bm{\Gamma}\right) \cdot \bm{\mu}(t) \cdot \mathbf{C}(t) + \int_0^t d\tau \mathbf{M_F}(t-\tau) \cdot \mathbf{C}(\tau) + \frac{1}{T}\left(\mathds{1} + \bm{\Gamma} \right) \cdot \mathbf{C}_0(p,t)  \label{eq:C} \\
        \begin{split}\label{eq:pspin-mu_F-TD-TTI}
        \bm{\mu}(t) &= \text{Diag}\Biggl[T\mathds{1}_2 + \int_0^t d\tau \mathbf{M}(t - \tau) \cdot \mathbf{C^T}(t - \tau) + \int_{0}^{t} d\tau \mathbf{D}(t - \tau) \cdot \partial_\tau\mathbf{F^T}(t - \tau) \\
        &+ \frac{1}{T}\left(\mathds{1} + \bm{\Gamma} \right) \cdot \mathbf{C}_0(p,t) \cdot \mathbf{C^T}(t)\Biggr] 
        \end{split}
\end{align}
\end{widetext}
where the Gaussian noise has correlations $\langle \bxi(t)\otimes\bxi(t')^T\rangle=2T{\bf 1}_2\delta(t-t')+{\bf D}(t,t')$. The three kernels ${\bf M_F}$, ${\bf C}_0$ and ${\bf D}$ that appear in Eqs.~\eqrefs{eq:effestocheq,eq:F,eq:C} are nonlinear functions of the elements of $\mathbf C$ and $\mathbf{F}$. The overall structure of these equations is similar to the equilibrium setting, though their derivation is somewhat more involved and we have deferred it to App.~\ref{app:pspindynamics}. These equations are solved with the boundary conditions $\mathbf{C}(0) = \mathds{1}$, $ \mathbf{F}(0) = \mathbf{0}$, while the explicit expression for the diagonal matrix $\bmu(t)$ enforces the spherical constraint.

In the case of standard equilibrium dynamics, the response and the correlations are related by the fluctuation-dissipation theorem (FDT), $\partial_t\mathbf{F}(t) = -\frac{1}{T}\partial_t\mathbf{C}(t)$ and it is possible to obtain a closed equation for the spin-spin correlation function of the system alone. By focusing on the long time limit of the dynamics one can then find the temperature below which the correlation function displays a nonzero plateau at infinite times when ergodicity is broken. Since the IO dynamics is a nonequilibrium one, we cannot resort to the fluctuation dissipation theorem, and analytical manipulations of~\eqrefs{eq:effestocheq,eq:F,eq:C,eq:pspin-mu_F-TD-TTI} may seem, at first sight, out of reach. In the following section, however, we will exhibit a simple argument that suggests that for the specific correlations and integrated response considered above, a relation formally similar to the FDT might still survive in the nonequilibrium steady-state.

\subsection{Response and fluctuations}
\subsubsection{Motivations} \label{sec:FDTHO}
The entropy production is a standard way of measuring the distance to equilibrium. Here we find, using Eq.~\eqref{eq:entropyprod}, that
\begin{equation}
\dot{\Sigma}=2\gamma^2 N\mu    
\end{equation}
so that the time scale at which nonequilibrium effects come into play goes as $\frac{1}{\gamma^2}$. It is tempting to believe that the distance to equilibrium can generically be viewed as being of order $O(\gamma^2)$. This may be true, but the connection between the response and the correlation functions, in our particular systems, runs almost as deep as in equilibrium. 

Before we address this question, it is useful, as a preliminary exercise, to work out the example of a harmonic oscillator undergoing the IO dynamics. This is a bit of an artificial example, as coarse-graining out one the two oscillators leads to an effective equilibrium dynamics for the remaining one (with memory, as discussed in Sec.~\ref{sec:heuristic}). Consider the two coupled equations for $\br=(x_1,x_2)$
\begin{equation}
    \dot{\br }=-(\mathds{1}+{\mathbf \Gamma})k\br +\sqrt{2T}\beta
\end{equation}
The evolution operator has a spectrum of relaxation rates $-k(n_1+n_2+i\gamma(n_1-n_2))$ (with $n_1,n_2\in\mathbb{N}$) and we realize that harmonic oscillators are marginal with respect to the acceleration of their dynamics: the relaxation rates only pick up a nonzero imaginary part expressing an overall rotation in phase space, but no modification of the real part means net acceleration. This system of two coupled oscillators is however indeed out of equilibrium since $\dot{\Sigma}=2 k\gamma^2>0$. The correlation matrix $\mathbf{C}(t,t')=\langle\br(t)\otimes\br(t')\rangle$ and the response function ${\mathbf R}(t,t')=\left.\frac{\delta\langle \br(t)\rangle}{\delta {\bf f}(t')}\right|_{\bf=\bo}$ to an external perturbing force $\mathbf{f}$ are related, in the stationary state, by 
\begin{equation}\label{eq:FDTHO}
    \mathbf R(t,t')(\mathds{1}+\bm\Gamma)=-\frac{1}{T}\p_t {\mathbf C}(t,t')
\end{equation}
which can be cast in an integral form by introducing the integrated response $\mathbf F(t,t') \equiv -\int_{t'}^t \dd\tau \mathbf{R}(t,\tau)$ as:
\begin{equation}\label{eq:FDTHOF}
    \mathbf F(t,t')(\mathds{1}+\bm\Gamma)=\frac{1}{T} \left( {\mathbf C}(t,t') - \mathbf C(t',t') \right)
\end{equation}
The fluctuation-dissipation theorem is not fulfilled, not even in some modified form, but it is true that the specific position-position correlation function is directly related to the integrated response in a linear fashion. There is no such generic connection for arbitrary observables. For lack of a better name, we christen this connection between these specific response and the correlations as an accidental FDT (aFDT). We shall now turn to a numerical investigation of the connection between $\mathbf F$ and $\mathbf C$ for the $p$-spin. We keep the {\it a priori} unrelated result of Eq.~\eqref{eq:FDTHO} under our belt for the analysis of the numerics which we carry out in the next subsection. 

\subsubsection{Numerics and aFDT}\label{sec:numerics}

\begin{figure}
    \centering
    \includegraphics[width = 0.4\textwidth]{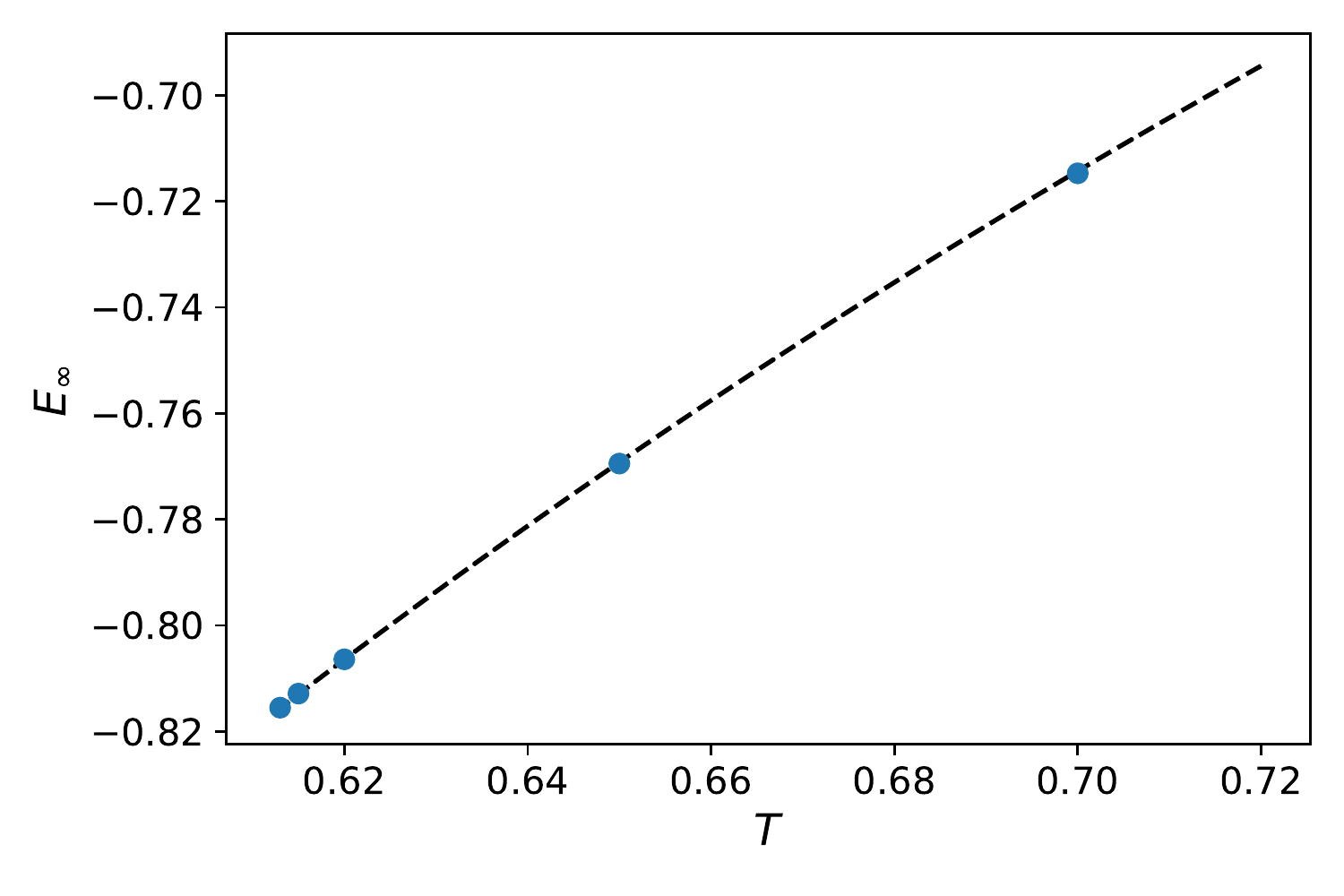}
    \caption{Energy per spin as a function of temperature $T$ obtained from numerical integration (blue dots) at $\gamma=1$ and theoretical prediction $E_{\infty} =-\frac{1}{2T}$ (black dashed line). }
    \label{fig:energy}
\end{figure}

We integrate the dynamical equations~\eqrefs{eq:F, eq:C, eq:pspin-mu_F-TD-TTI} for $p=3$ using a finite difference method implemented in \cite{fuchs1991comments, kim2001dynamics, berthier2007spontaneous, folena2020mixed} and detailed in App.~\ref{app:numerics}. We check that our implementation of the IO dynamics samples the correct Boltzmann distribution by measuring the energy per spin $E_\infty$ at long times in the ergodic phase. This is a static quantity independent from $\gamma$, for which a value of $E_{\infty} = -\frac{1}{2T}$ is predicted~\cite{crisanti1992spherical}. In the numerical integration, we computed the energy by exploiting the relation
\begin{equation}
    \bm{\mu}(t \rightarrow +\infty) =  \left(T - pE_{\infty}\right)\mathds{1}
\end{equation}
The agreement between the measured value of $E_{\infty}$ and its theoretical prediction is shown in Fig.~\ref{fig:energy}. The errors are of order $10^{-12}$ and they are therefore  negligible.  

\begin{figure}
    \centering
    \includegraphics[width = 0.4\textwidth]{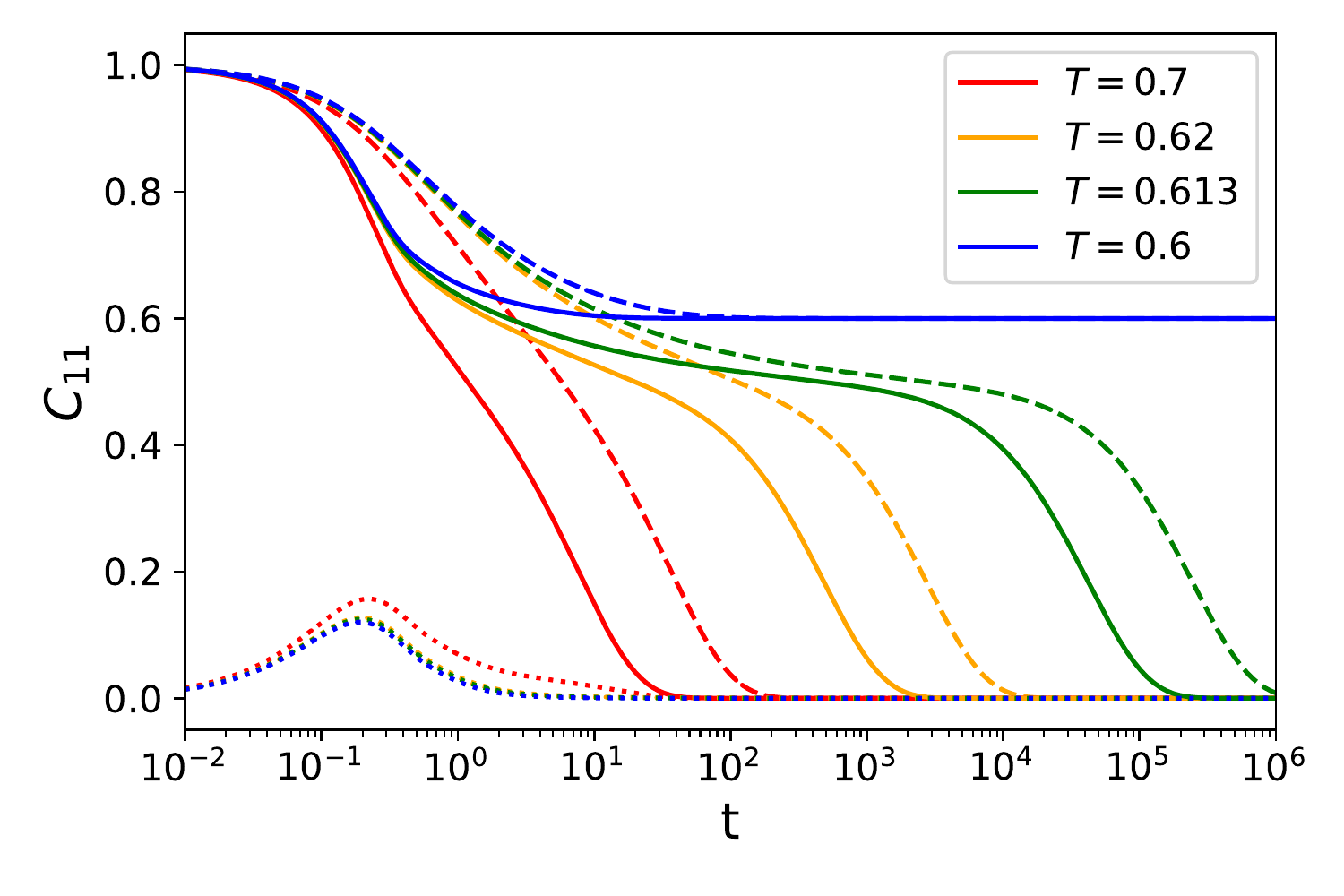}
    \caption{Correlation function of system $1$, $C_{11}(t)$ for $\gamma=2.4$ (solid lines) and $\gamma = 0$ (dashed lines) for various temperatures above and below $T_d \approx 0.6124$. The dotted curves represent the cross correlations $C_{12}(t)$ for $\gamma=2.4$. The shift between the dashed and the solid line of a given color, which quantifies the acceleration of the dynamics, somewhat increases as the temperature is decreased.}
    \label{fig:C11}
\end{figure}

Having ascertained that our implementation of the IO dynamics correctly samples a static quantity, we proceed by investigating the time behavior of the correlation function of one subsystem. Our results are presented in Fig.~\ref{fig:C11}. In the range of parameters explored, with $\gamma$ up to $2.4$, the IO dynamics is accelerated by a non negligible amount compared to the standard one. For $\gamma \approx 2.4$, the highest coupling achievable while preserving the convergence of the integration scheme, the convergence to $0$ of the correlation function is anticipated by almost an order of magnitude. For $T < T_d$, the system freezes even at $\gamma \neq 0$, and the position of the infinite time plateau is left unchanged. 

\begin{figure}
    \centering
    \includegraphics[width = 0.4\textwidth]{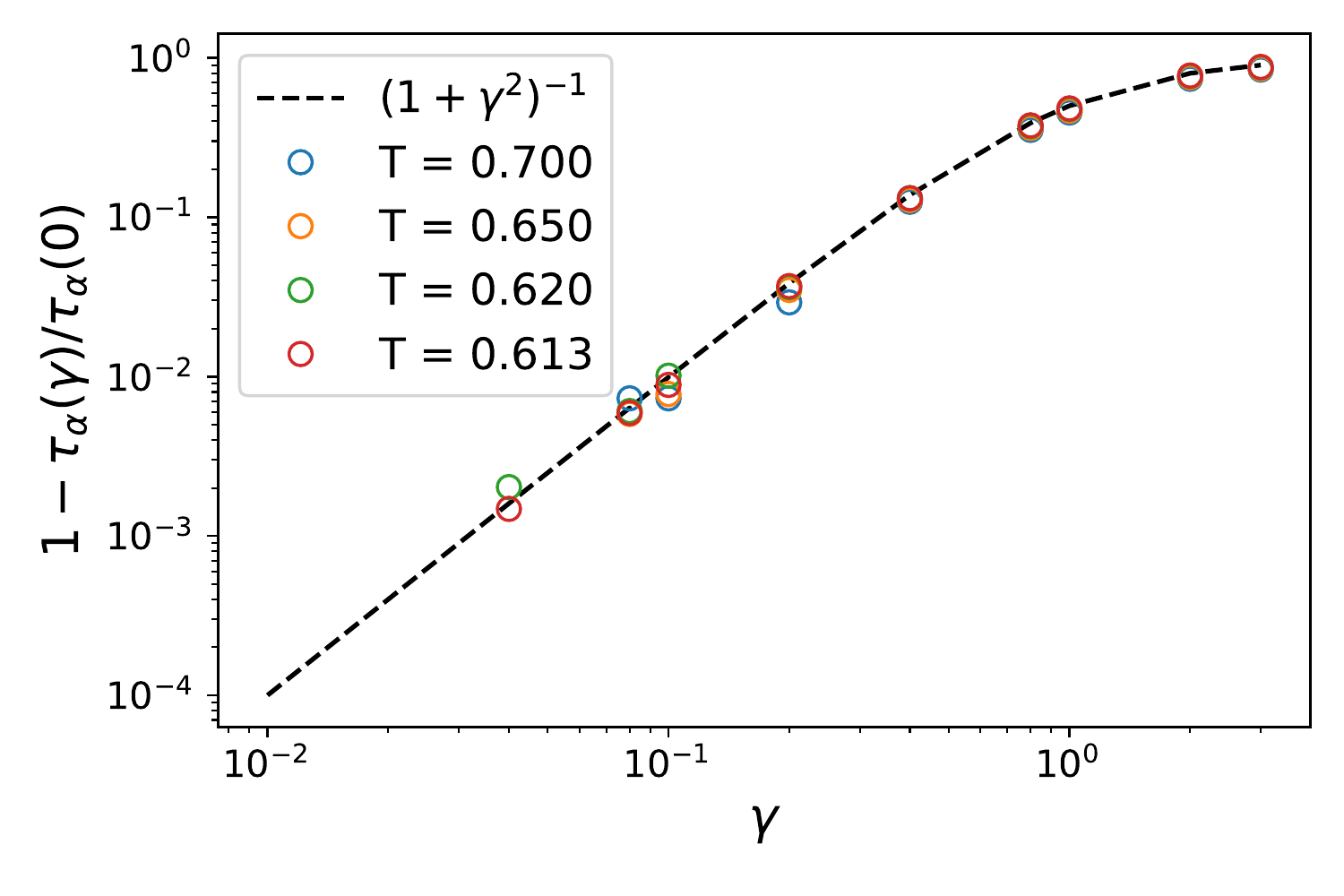}
    \caption{Ratio between the relaxation time of $C_{11}$ for the IO dynamics, $\tau_{\alpha}(\gamma)$, and the equilibrium dynamics $\tau_{\alpha}(0)$ for temperatures close to $T_d$.  }
    \label{fig:taualpha}
\end{figure}

We quantify the speedup obtained by means of the IO dynamics by measuring the relaxation time $\tau_\alpha(\gamma)$, defined here as the time necessary for $C_{11}$ to decay to the value $1/e$ in the ergodic phase. Our results are shown in Fig.~\ref{fig:taualpha}. The ratio  between $\tau_\alpha(\gamma)$ and the relaxation time in the standard dynamics, $\tau_{\alpha}(0)$ appears to scale approximatily as $(1+\gamma^2)^{-1}$. This scaling the same as the one found in Sec. \ref{sec:heuristic} when heuristically  discussing the Kramers' problem. We will provide the reader with further interpretation for this phenomenon in the next section.

\begin{figure}
     \centering
     \includegraphics[width = 0.4\textwidth]{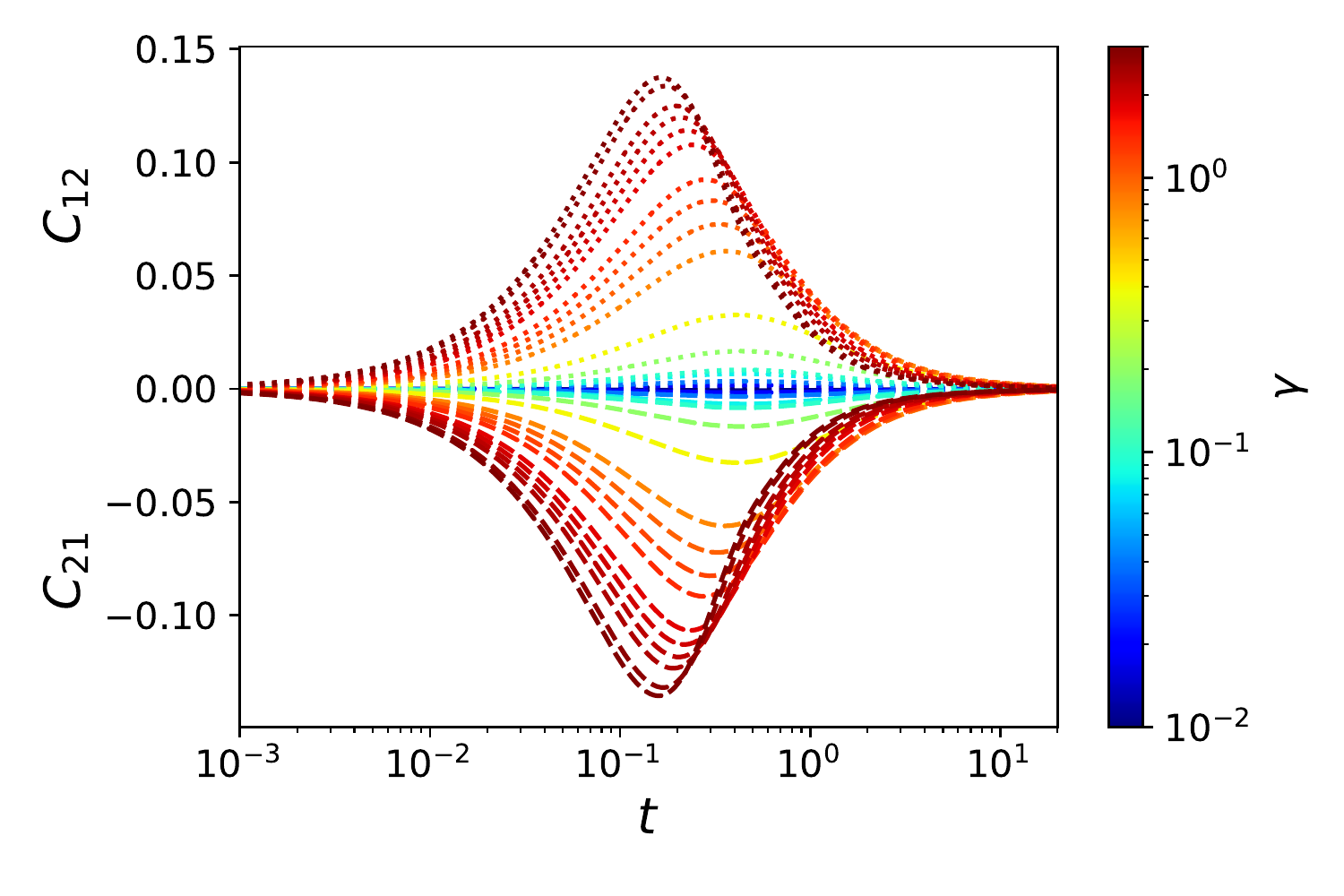}
     \caption{Cross correlation $C_{12}$ (dotted lines) and $C_{21}$ (dashed lines)as a function of time for $T=0.613$ at different values of $\gamma$.}
     \label{fig:C12}
 \end{figure}

 The departure of the IO dynamics from the equilibrium dynamics is accompanied by an early growth in the cross correlations of the system. This can be seen in Fig.~\ref{fig:C11}, where we plot $C_{12}$ in dotted lines, and is presented in more detail in Fig.~\ref{fig:C12}, where a close up of the cross correlations for various values of $\gamma$ is shown. The dynamics of these quantities is entirely located in the region of $\beta$-relaxation of $C_{11}$. They exhibit a bump at short times, reaching a point of maximum modulus and then decaying to zero. This maximum grows linearly with $\gamma$ for $\gamma \lesssim 1$. The location of the maximum moves left  to shorter times as $\gamma$ grows, as expected from Eq.~\eqref{eq:entropyprod}. 
 
\begin{figure}
    \centering
    \includegraphics[width = 0.4\textwidth]{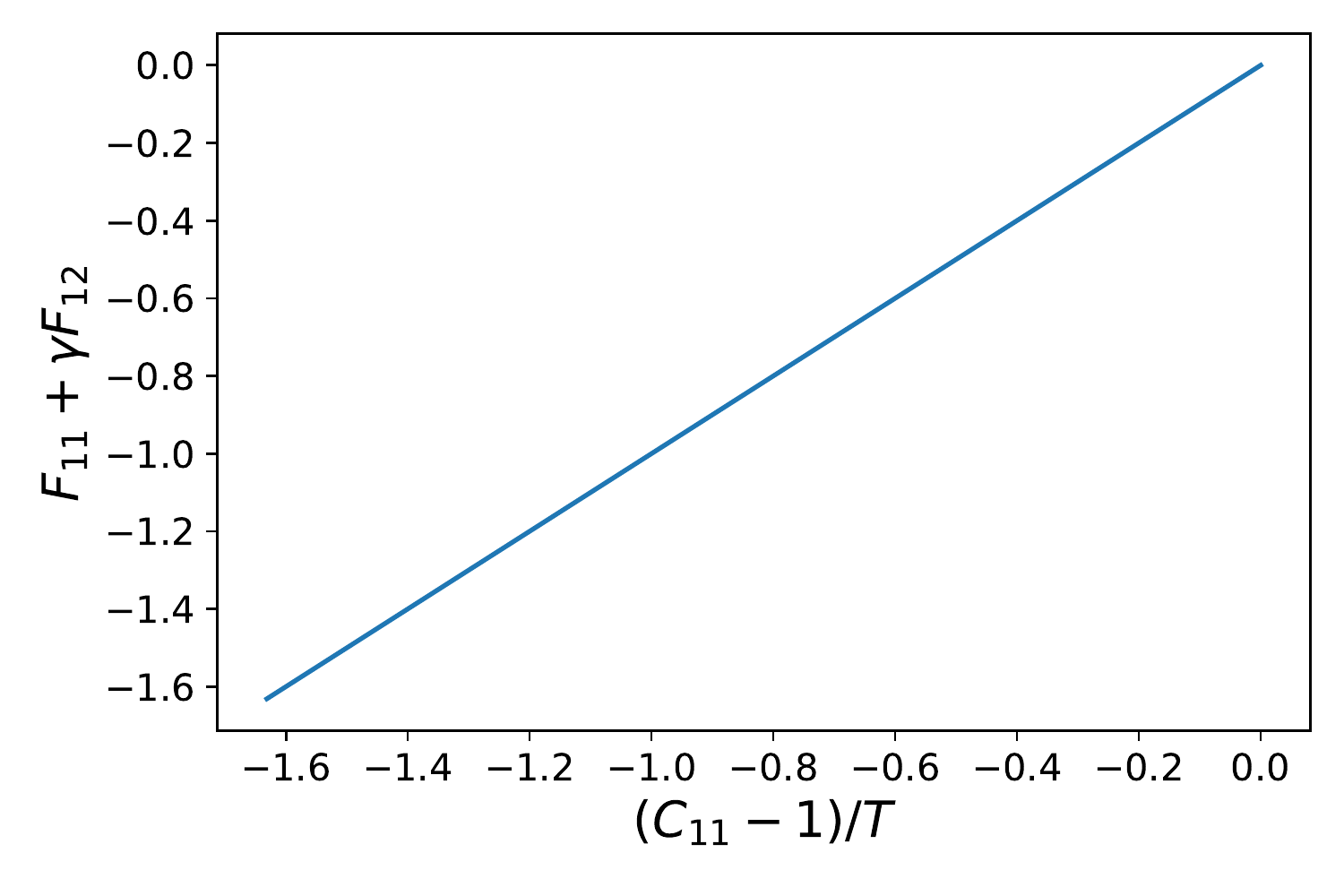}
    \caption{Parametric plot of the left and right hand side of the first entry of Eq.~\eqref{eq:FDTHOF} from the integration of ~\eqrefs{eq:F, eq:C, eq:pspin-mu_F-TD-TTI} at $T=0.613$ and $\gamma = 2.4$.}
    \label{fig:FDTHO}
\end{figure}

 Guided by the analysis of the harmonic oscillator in Sec.~\ref{sec:FDTHO}, we investigate whether Eq.~\eqref{eq:FDTHO} holds for the $p$-spin, by studying the relationship between each entry of the correlation matrix and the right hand side of Eq.~\eqref{eq:FDTHOF} and integrated response matrices. An example for a particular entry and particular values of $T$ and $\gamma$ is shown in Fig.~\ref{fig:FDTHO}. There the equality of Eq.~\eqref{eq:FDTHOF} previously found for the harmonic oscillator is clearly verified. Similar results are obtained for other entries and other values of coupling and temperature. This is a consequence of the effective linear dynamics of $\bsigma$ in Eq.~\eqref{eq:effestocheq}.

The numerical investigation performed in this section raises a series of questions to be answered. Can we show that the aFDT actually holds in the $p$-spin IO dynamics? Can this result explain the absence of a shift in the dynamical temperature $T_d$? Is it possible to justify the scaling of the relaxation time of $C_{11}$ with $\gamma$?  In the following section we will build a theoretical understanding of our numerical findings.

\subsubsection{Interpretation of the aFDT}
We establish the validity of the aFDT relation for the IO dynamics of the $p-$spin. We can indeed show that if ~\eqrefs{eq:C, eq:pspin-mu_F-TD-TTI} and the aFDT relation Eq. \eqref{eq:FDTHO} are true then the remaining equation Eq.~\eqref{eq:F} must also hold. The details of the proof are reported in App.~\ref{app:FDTHO}. 

By using the aFDT relation we can reduce the sets of ~\eqrefs{eq:F, eq:C, eq:pspin-mu_F-TD-TTI} to a single equation for the correlation matrix $\mathbf{C}(t)$:
\begin{equation}\label{eq:CFDTHO}
\begin{split}
    \p_t\mathbf{C}(t) &= -T\left(\mathds{1} + \bm \Gamma \right) \cdot \mathbf{C}(t) \\
    &- \frac{1}{T}\left(\mathds{1} + \bm \Gamma \right) \int_0^{t}d\tau \mathbf{C}_0(p,t-\tau) \cdot \partial_\tau \mathbf{C}(\tau)
\end{split}
\end{equation}
with $C_0(p,t)_{ij} \equiv \frac{p}{2}C_{ij}^{p-1}$. Apart from the matrix prefactor $\left(\mathds{1} + \bm \Gamma \right)$  the structure of this equation is the same as the one obtained in the standard equilibrium dynamics of the $p$-spin \cite{crisanti1993sphericalp, barrat1997p, castellani2005spin}. We therefore study its long time behavior by assuming that $\lim_{t \rightarrow +\infty} \mathbf{C}(t) = \mathbf{Q}$, with $\mathbf{Q}$  a matrix of long time plateaus generalizing the nonergodic parameter. We introduce the Laplace transform of a matrix $\mathbf{A}(t)$ as $\mathbf{\tilde{A}}(z) \equiv \int_0^{+\infty} \dd t \mathbf{A}(t)\ee^{-zt}$ and we observe that $\mathbf{\tilde{C}}(z) \approx \frac{\mathbf{Q}}{z}$ and $\mathbf{\tilde{C}}_0(p,z) \approx \frac{\mathbf{Q}_0(p)}{z}$ as $z \rightarrow 0$, with $Q_0(p)_{ij} \equiv Q_{ij}^{p-1}$. Applying the Laplace transform to both sides of Eq.~\eqref{eq:CFDTHO} and keeping only the terms that diverge as $z\to 0$ yields
\begin{equation}\label{eq:CFDTHOLaplace}
    T\mathbf{Q} - \frac{1}{T}\mathbf{Q}_0(p) \cdot \left(\mathds{1} - \mathbf{Q} \right) = 0
\end{equation}
We now see that the dependence on $\mathbf{\Gamma}$ has simply disappeared. The form of the plateaus matrix can therefore not depend on $\gamma$. In particular, it must match the solution obtained at $\gamma = 0$. We can therefore write $\mathbf{Q} = Q\mathds{1}$ and $\mathbf{Q}_0(p) = Q^{p-1}\mathds{1}$. Eq.~\eqref{eq:CFDTHOLaplace} then becomes the standard equation for the long time plateau of the $p$-spin equilibrium dynamics
\begin{equation}
    \frac{2T^2}{p} = Q^{p-2}(1 - Q)
\end{equation}
In analogy with the Kramers problem and the harmonic oscillator of Sec.~\ref{sec:kramers} we see that an  increased mobility $\left(\mathds{1} + \bm \Gamma \right)$ arises in the IO dynamics of the $p$-spin. This rescaling of the mobility can always be absorbed into a rescaling of the time units. Below $T_d$, where the plateau extends to infinite times, the dynamical ergodicity breaking cannot be avoided while above $T_d$ an acceleration is achieved.

We support the above picture by carrying out a perturbation expansion to second order in $\gamma$. In particular we will recover the scaling of the relaxation time $\tau_{\alpha}$ shown in Fig.~\ref{fig:taualpha} starting from the dynamical equation for the correlation matrix Eq.~\eqref{eq:CFDTHO}. We multiply both sides by $\mathds{1} + \bm{\Gamma^T}$ and we expand $\mathbf{C}$ in powers of $\gamma$, $\mathbf{C} \approx \delta_0 \mathbf{C} + \gamma\delta_1\mathbf{C} + \gamma^2\delta_2\mathbf{C}$, thereby obtaining:
\begin{equation}\label{eq:perturbationCFDTHO}
    \begin{split}
        \partial_t\mathbf{C}(t) + \bm{\Gamma^T}\cdot \partial_t \delta_1 \mathbf{C}(t) = &-(1+\gamma^2)\Biggl(\mathbf{C}(t) \\
        &+ \frac{1}{T}\int_0^t d\tau \mathbf{C}_0(p,t-\tau) \cdot \p_\tau\mathbf{C}(\tau)\Biggr)
    \end{split}
\end{equation}
The second term on the left hand side contains $\delta_1\mathbf{C}$, the first order perturbation of $\mathbf{C}$. It is a skew symmetric matrix and therefore it contributes to the expansion of the cross correlations shown in Fig.~\ref{fig:C12}. Motivated by the numerical observation of Sec.~\ref{sec:numerics} we assume that this term essentially vanishes at times larger than the time scale at which the $\beta$ relaxation takes place. Moreover, because of the skew symmetry of the first order perturbation, the diagonal entries of Eq.~\eqref{eq:perturbationCFDTHO} are not affected by the short time dynamics of $\delta_1 \mathbf{C}$. We therefore obtain, to second order in $\gamma$:
\begin{equation}
    \begin{split}
        \partial_t C_{aa}(t) = &-(1+\gamma^2)\Biggl(TC_{aa}(t) \\
        &+ \frac{1}{T}\int_0^t d\tau C_{aa,0}(p,t-\tau) \p_\tau C_{aa}(\tau)\Biggr)
    \end{split}
\end{equation}
for $a=1,2$. This is the same equation as that of the equilibrium dynamics, up to a rescaling of time $t \rightarrow t/(1 + \gamma^2)$, in agreement with the findings of Fig.~\ref{fig:taualpha}.

\section{Prospects}
The nonequilibrium dynamics investigated here accelerates the relaxation of the $p$-spin model without shifting the temperature at which a dynamical ergodicity breaking occurs. A quantitative estimate of the speed up of the dynamics is that an overall factor $1+\gamma^2$ in time is gained, for not too large values of $\gamma$. We have come up with an explanation for this, building on the analysis of the Kramers' time for the IO dynamics. The dynamics effectively yields an enhanced mobility, which accelerates the dynamical evolution without affecting the effective height of the energetic barrier. The presence of enhanced mobility explains both the preservation of $T_d$ and the acceleration in the ergodic phase. We have provided evidence for the scaling form of the acceleration with $\gamma$. At finite number $N$ of spins, it would interesting to probe the interplay of $\gamma$ with the system size, as in principle, a sufficiently large $\gamma$ could overcome energy barriers. To adjust $\gamma$ appropriately, one could rely on the analysis of Stariolo and Cugliandolo~\cite{stariolo2019activated,stariolo2020barriers}, where the relevant time-scales are carefully sorted out. A related interesting question is how the instanton trajectories are modified by a nonzero $\gamma$ in a finite size mean-field model, expanding on the analysis of the Kramers problem made in the present article. A relevant comparison could be provided by the study of Ros \textit{et al.}~\cite{ros2021dynamical} where the instanton trajectory in the spherical $p$-spin model under equilibrium overdamped Langevin dynamics is characterized. 

While the numerical scheme we resort to prohibits the investigation of  values of $\gamma$ beyond $O(1)$, it is unclear at which value of $\gamma$  the saturation of the speed up will occur. However, we speculate that such a saturation is difficult to avoid. An interpretation of this effect that we predict can be connected to the simple harmonic oscillator. Indeed, as $\gamma$ is increased, one system is injected with a larger amount of energy by the other one, but most of the time is spent swirling around in a circle a finite distance from the bottom of the well, and this is exactly what would happen in a harmonic oscillator, for which no acceleration is ever achieved. We believe that coupling $n$ systems together would also lead to a marginal gain as a similar saturation effect in terms of $\gamma_\text{eff}\sim \sqrt{n}\gamma$ would occur for too large $n$. This, in turn, raises the question of how to fine tune the optimal value of $\gamma$. As our results suggest, it is possible that the best acceleration is achieved by a nontrivial time-dependent protocol $\gamma(t)$ (in which $\gamma(\to+\infty)\to\gamma_\infty$, the latter being possibly zero). This question is touched upon in \cite{lelievre2013optimal}.

Beyond these formal but interesting questions, what we believe to be of true physical interest is the application of such dynamics to molecular glass formers. Unlike the $p$-spin, realistic glasses display only finite energy barriers and the effect of $\gamma$ should be felt at all temperatures. How this approach fares as compared to other ones such as SWAP~\cite{grigera2001fast, berthier2016equilibrium, ninarello2017models} is, in our opinion, and based on preliminary results~\cite{prel}, the most interesting open question.
\begin{acknowledgments}
We acknowledge very insightful exchanges with A. Altieri, T. Arnoulx de Pirey, L. Berthier, L.F. Cugliandolo, J. Tailleur. We acknowledge the financial support of the ANR grant THEMA.
\end{acknowledgments}

\appendix

\onecolumngrid

\section{Instanton trajectory close to the saddle point} \label{app:instantonlinearised}
We study the instanton trajectory in the vicinity of the saddle point $\mathbf{x}_s$. The first line of Eq.~\eqref{eq:instanton} reads
\begin{equation}\label{eq:IOinstantonlinear}
    \dot{\mathbf{x}} = \mathbf{A} \cdot \mathbf{x}
\end{equation}
with 
\begin{equation}
    \mathbf{A} \equiv \begin{bmatrix}
                    -k_M & \gamma k_m \\
                    \gamma k_M & k_m
                \end{bmatrix}
\end{equation}
This linear systems has one unstable and one stable direction along which the dynamics takes the form $C_\pm e^{\lambda_{\pm}t}\mathbf{v}_\pm$,with $\lambda_{\pm}$, $\mathbf{v}_{\pm}$ the eigenvalues and eigenvectors of $\mathbf{A}$ and $C_\pm$ determined by the initial condition. Both the eigenvalues and the eigenvectors depend on $\gamma$. The $t\rightarrow +\infty$ limit of the instanton dynamics in the first line of Eq.~\eqref{eq:IOinstantonlinear} corresponds to the dynamics of system  along its stable direction $\mathbf{v}_-$. By expanding the latter around $\gamma = 0$ we obtain
\begin{equation}
    e^{\lambda_- t}\mathbf{v}_- =   \begin{bmatrix}
                                        -\ee^{-k_M t} + \gamma^2 \ee^{-k_M t}\frac{k_M^2 + 2k_mt(k_m + k_M)}{2(k_m + k_M)^2} + {O}(\gamma^4) \\
                                        \gamma \ee^{-k_M t}\frac{k_M}{k_m + k_M} + {O}(\gamma^3)
                                    \end{bmatrix}
\end{equation}      
which justifies the form of the perturbation expansion adopted in Eq.~\eqref{eq:perturbationtheory}.

\section{Dynamical equations for the $p$-spin evolving under the IO dynamics}\label{app:pspindynamics}

We consider two copies of the spherical $p$-spin model evolving according to the Ichiki-Ohzeki (IO) dynamics, for which the equations of motion read
\begin{align}\label{eq:Langevinpspinapp}
\p_t\sigma_i^{(1)}=-\frac{\p {\mathscr H}}{\p\sigma_i^{(1)}}+\gamma\frac{\p {\mathscr H}}{\p\sigma_i^{(2)}}- \mu^{(1)}(t)\sigma_i^{(1)} + \gamma \mu^{(2)}(t)\sigma_i^{(2)}+\sqrt{2T}\xi_i^{(1)}\\
\p_t\sigma_i^{(2)}=-\frac{\p {\mathscr H}}{\p\sigma_i^{(2)}}-\gamma\frac{\p {\mathscr H}}{\p\sigma_i^{(1)}}-\mu^{(2)}(t)\sigma_i^{(2)} - \gamma\mu^{(1)}(t)\sigma_i^{(1)}+\sqrt{2T}\xi_i^{(2)}
\end{align}
where $\mu^{(a)}(t)$ are elastic constants that enforce the spherical constraints $\sum_{i=1}^N \left\langle \left(\sigma_i^{(a)}\right)^2 (t)\right\rangle = N$ for $a=1,2$, $\xi_i^{(a)}$ are independent Gaussian white noises $\left\langle \xi_i^{(a)}(t)\xi_j^{(b)}(t') \right\rangle = \delta_{ab}\delta_{ij}\delta(t-t')$ and the total Hamiltonian of the system $\mathscr{H}$ reads
\begin{equation}
    \mathscr{H}\left[\{\sigma_i\}^{(a)}_{i=1,\ldots,N, a = 1,2}\right]=-\sum_{i_1<\ldots<i_p}J_{i_1\ldots i_p}\sum_{a=1,2}\sigma^{(a)}_{i_1}\ldots\sigma^{(a)}_{i_p}
\end{equation}
Note that the couplings $J_{i_1\ldots i_p}$ have the same realization for both systems, and they are independently distributed according to  a Gaussian distribution of mean $0$ and variance $\frac{p!}{2N^{p-1}}$.

We suppose that the systems start from an initial equilibrium condition at a temperature $T \equiv \beta^{-1}$, which is kept constant throughout the dynamical evolution. Our goal is to derive effective dynamical equations of motion for the spins in the two system by averaging over the disorder. To achieve this, we use standard path integral techniques as carried out  in \cite{cugliandolo1993analytical} and further detailed in \cite{barrat1997p} or \cite{castellani2005spin}. The starting point is the partition function for the dynamics of the composite system in Eq.~\eqref{eq:Langevinpspinapp} at temperature $T$. The initial condition for the two subsystems is given by two independent realization of the Boltzmann distribution at temperature $T$. We have therefore
\begin{equation}
    \begin{aligned}
    1 \equiv Z &=\int \mathcal{D}\bm{\sigma}_0\frac{1}{Z_{\text{eq}}}e^{-\beta\mathscr H[\sigma_0^{(1)},\sigma_0^{(2)}]} \\
    &\times \int \mathcal{D} \bm{\sigma} \mathcal{D} \bm{\hat\sigma} \exp \Biggl[- \sum_{i=1}^N \int_0^t \dd\tau \hat\sigma_i^{(1)}\left(\partial_\tau\sigma^{(1)}_i +\frac{\p {\mathscr H}}{\p\sigma_i^{(1)}}-\gamma\frac{\p {\mathscr H}}{\p\sigma_i^{(2)}}+ \mu^{(1)}\sigma_i^{(1)} - \gamma \mu^{(2)}\sigma_i^{(2)} \right) \\
    &+ \hat\sigma_i^{(2)}\left(\partial_\tau \sigma^{(2)}_i +\frac{\p {\mathscr H}}{\p\sigma_i^{(2)}}+\gamma\frac{\p {\mathscr H}}{\p\sigma_i^{(1)}} + \mu^{(2)}\sigma_i^{(2)} + \gamma \mu^{(1)}\sigma_i^{(2)}\right)
    - T\left(\hat{\sigma}_i^{(1)}\right)^2 - T\left(\hat{\sigma}_i^{(2)}\right)^2\Biggr]
\end{aligned}
\end{equation}
with $Z_{\text{eq}}$ the static partition function of the system, $Z_{\text{eq}} \equiv \int \mathcal{D} \bm{\sigma}_0 e^{-\beta \mathscr H[\bm{\sigma}_0]}$. We are using a vector notation for the spin variables, $\bm{\sigma}(t)=\left(\sigma^{(1)}(t),\sigma^{(2)}(t)\right)^T$, the auxiliary field $\bm{\hat\sigma}(t)=\left(\hat\sigma^{(1)}(t),\hat\sigma^{(2)}(t)\right)^T$ and the initial condition $\bm{\sigma}_0(t)=\left(\sigma^{(1)}(0),\sigma^{(2)}(0)\right)^T$ As noted in \cite{houghton1983role}, since we are using thermalized initial conditions, we must resort to the replica trick in order to average $Z$ over the disorder:
\begin{equation}
    \frac{1}{Z_\text{eq}} = \lim_{n\rightarrow 0} Z_{\text{eq}}^{n-1}
\end{equation}
and follow the dynamics of each replica. However, since the initial condition belongs to the replica symmetric phase, the dynamics of the different replicas are decoupled and it suffices to follow the behavior of any single one among them. By using notation of the form $\left(\sigma^{(a)}\cdot\sigma^{(b)}\right)(t,t') \equiv \sum_{i=1}^N \sigma^{(a)}_i(t)\sigma^{(b)}_i(t')$, with $a$ and $b$ indices identifying the two systems,  we can write in the thermodynamic limit 
\begin{equation}\label{eq:Zavg1}
    \begin{aligned}
    \overline{Z} &= \int \mathcal{D}\bm{\sigma}(0)\mathcal{D}\bm{\sigma}\mathcal{D}\bm{\hat\sigma} \exp \Biggl\{-\sum_{i=1}^N \Biggl[\int \dd\tau  \hat\sigma_i^{(1)}\left(\partial_\tau \sigma_i^{(1)} + \mu^{(1)}\sigma_i^{(1)} - \gamma\mu^{(2)}\sigma_i^{(2)} \right) - T\left(\hat\sigma_i^{(1)}\right)^2 \\
    & +  \hat\sigma_i^{(2)}\left(\partial_\tau \sigma_i^{(2)} + \mu^{(2)}\sigma_i^{(2)} + \gamma\mu^{(1)}\sigma_i^{(1)} \right) - T\left(\hat\sigma_i^{(2)}\right)^2\Biggr] + \frac{1}{4N^{p-1}}\int \dd t\int  \dd t' I_1(t,t') + \frac{1}{4TN^{p-1}}\int \dd t I_2(t)\Biggr\}
    \end{aligned}
\end{equation}
With
\begin{equation}
    \begin{aligned}
    I_1(t,t') &\equiv p\left(\hat\sigma^{(1)}\cdot \hat\sigma^{(1)}\right)\left(\sigma^{(1)}\cdot \sigma^{(1)} \right)^{p-1} + p(p-1)\left(i\hat\sigma^{(1)} \cdot \sigma^{(1)}\right)\left(\sigma^{(1)} \cdot \hat\sigma^{(1)}\right)\left(\sigma^{(1)} \cdot \sigma^{(1)} \right)^{p-2} \\
    &-\gamma \left[ p\left(\hat\sigma^{(1)}\cdot \hat\sigma^{(1)}\right)\left(\sigma^{(1)}\cdot \sigma^{(2)} \right)^{p-1} + p(p-1)\left(\hat\sigma^{(1)} \cdot \sigma^{(2)}\right)\left(\sigma^{(1)} \cdot\hat\sigma^{(1)}\right)\left(\sigma^{(1)} \cdot \sigma^{(2)} \right)^{p-2} \right] \\
    &+ p\left(\hat\sigma^{(1)}\cdot \hat\sigma^{(2)}\right)\left(\sigma^{(1)}\cdot \sigma^{(2)} \right)^{p-1} + p(p-1)\left(\hat\sigma^{(1)} \cdot \sigma^{(2)}\right)\left(\sigma^{(1)} \cdot \hat\sigma^{(2)}\right)\left(\sigma^{(1)} \cdot \sigma^{(2)} \right)^{p-2} \\
    &+\gamma \left[p\left(\hat\sigma^{(1)}\cdot \hat\sigma^{(2)}\right)\left(\sigma^{(1)}\cdot \sigma^{(1)} \right)^{p-1} + p(p-1)\left(\hat\sigma^{(1)} \cdot \sigma^{(1)}\right)\left(\sigma^{(1)} \cdot \hat\sigma^{(2)}\right)\left(\sigma^{(1)} \cdot \sigma^{(1)} \right)^{p-2} \right] \\
    &-\gamma \left[ p\left(\hat\sigma^{(1)}\cdot \hat\sigma^{(1)}\right)\left(\sigma^{(2)}\cdot \sigma^{(1)} \right)^{p-1} + p(p-1)\left(\hat\sigma^{(1)} \cdot \sigma^{(1)}\right)\left(\sigma^{(2)} \cdot \hat\sigma^{(1)}\right)\left(\sigma^{(2)} \cdot \sigma^{(1)} \right)^{p-2}\right]\\
    &+\gamma^2 \left[p\left(\hat\sigma^{(1)}\cdot \hat\sigma^{(1)}\right)\left(\sigma^{(2)}\cdot \sigma^{(2)} \right)^{p-1} + p(p-1)\left(\hat\sigma^{(1)} \cdot \sigma^{(2)}\right)\left(\sigma^{(2)} \cdot \hat\sigma^{(1)}\right)\left(\sigma^{(2)} \cdot \sigma^{(2)} \right)^{p-2} \right]\\
    &-\gamma \left[p\left(\hat\sigma^{(1)}\cdot \hat\sigma^{(2)}\right)\left(\sigma^{(2)}\cdot \sigma^{(2)} \right)^{p-1} + p(p-1)\left(\hat\sigma^{(1)} \cdot \sigma^{(2)}\right)\left(\sigma^{(2)} \cdot \hat\sigma^{(2)}\right)\left(\sigma^{(2)} \cdot \sigma^{(2)} \right)^{p-2} \right] \\
    &-\gamma^2 \left[p\left(\hat\sigma^{(1)}\cdot \hat\sigma^{(2)}\right)\left(\sigma^{(2)}\cdot \sigma^{(1)} \right)^{p-1} + p(p-1)\left(\hat\sigma^{(1)} \cdot \sigma^{(1)}\right)\left(\sigma^{(2)} \cdot \hat\sigma^{(2)}\right)\left(\sigma^{(2)} \cdot \sigma^{(1)} \right)^{p-2} \right] \\
    & + \left(1 \leftrightarrow 2, \gamma \leftrightarrow -\gamma\right)
    \end{aligned}
\end{equation}
and
\begin{equation}
    \begin{aligned}
    \frac{1}{2}I_2(t,0) &\equiv p\left(\hat\sigma^{(1)} \cdot \sigma_0^{(1)} \right)\left( \sigma^{(1)}\cdot \sigma_0^{(1)} \right)^{p-1} - \gamma p \left(\hat\sigma^{(1)} \cdot \sigma^{(1)}_0 \right)\left(\sigma^{(2)}\cdot \sigma_0^{(1)} \right)^{p-1} \\
    &+ p\left(\hat\sigma^{(1)} \cdot \sigma_0^{(2)} \right)\left( \sigma^{(1)}\cdot \sigma_0^{(2)} \right)^{p-1} - \gamma p \left(\hat\sigma^{(1)} \cdot \sigma^{(2)}_0 \right)\left(\sigma^{(2)}\cdot \sigma_0^{(2)} \right)^{p-1} \\
    &+ \left(1 \leftrightarrow 2, \gamma \leftrightarrow -\gamma\right)
    \end{aligned}
\end{equation}
We now exploit the thermodynamic limit $N \rightarrow \infty$ through a saddle point evaluation of the path integral. In order to do so, we introduce a set of dynamical overlaps in the partition function:
\begin{equation} \label{eq:Zavgdelta}
    \begin{aligned}
    \overline{Z} &= \int \mathcal{D}\mathbf{Q} \Pi_{a,b = 1,2}\delta\left(NQ_1^{(ab)}(t,t') - \hat\sigma^{(a)}\cdot \hat\sigma^{(b)}\right)\delta\left(NQ_2^{(ab)}(t,t') - \sigma^{(a)}\cdot\sigma^{(b)}\right) \\
    &\times\delta\left(NQ_3^{(ab)}(t,t') - \hat\sigma^{(a)}\cdot\sigma^{(b)}\right)\delta\left(NQ_4^{(ab)}(t,t') - \sigma^{(a)}\cdot \hat\sigma^{(b)}\right) \\
    &\times \delta\left(NQ_5^{(ab)}(t) - \hat\sigma^{(a)} \cdot \sigma_0^{(b)} \right)\times \delta\left(NQ_6^{(ab)}(t) - \sigma^{(a)} \cdot \sigma_0^{(b)} \right)\times \left(\text{r.h.s. Eq. \eqref{eq:Zavg1}}\right) \\
    &= \int\mathcal{D}\mathbf{P}\int\mathcal{D}\mathbf{Q} \Pi_{a,b=1,2}\exp \left[iN \int \dd t \dd t'\left(P_1^{(ab)}(t,t')Q_1^{(ab)} - P_1^{(ab)}(t,t')\hat\sigma^{(a)}\cdot \hat\sigma^{(b)} \right)  \right] \\
    &\times \exp \left[iN \int \dd t \dd t'\left(P_2^{(ab)}Q_2^{(ab)} - P_2^{(ab)}\sigma^{(a)}\cdot\sigma^{(b)} \right)  \right] \times \exp \left[iN \int \dd t \dd t'\left(P_3^{(ab)}(t,t')Q_3^{(ab)} - P_3^{(ab)} \hat\sigma^{(a)}\cdot\sigma^{(b)} \right)  \right] \\
    &\times \exp \left[iN \int \dd t \dd t'\left(P_4^{(ab)}Q_4^{(ab)} - P_4^{(ab)}\sigma^{(a)}\cdot \hat\sigma^{(b)} \right)  \right] \times \exp \left[iN \int \dd t \left(P_5^{(ab)}Q_5^{(ab)} - P_5^{(ab)} \hat\sigma^{(a)}\cdot \sigma_0^{(b)} \right)  \right] \\
    &\times \exp \left[iN \int \dd t \left(P_6^{(ab)}Q_6^{(ab)} - P_6^{(ab)}\sigma^{(a)}\cdot\sigma_0^{(b)} \right)  \right] \times \left(\text{r.h.s of Eq. \eqref{eq:Zavg1}}\right) 
    \end{aligned}
\end{equation}
The dynamical overlaps have a natural interpretation if one introduces the correlation and the response matrices $\mathbf{C}(t,t')$ and $\mathbf{R}(t,t')$:
\begin{eqnarray}
    C_{ab}(t,t') &\equiv \frac{1}{N}\sum_{i}\langle \sigma_i^{(a)}(t)\sigma_i^{(b)}(t')\rangle \label{eq:corr}\\
    R_{ab}(t,t') &\equiv \frac{1}{N}\sum_{i}\langle \sigma_i^{(a)}(t)\hat\sigma_i^{(b)}(t') \rangle
\end{eqnarray}
We therefore see that in the thermodynamic limit the following self consistent relations hold:
\begin{equation} \label{eq:intdynpar}
\begin{cases}
Q_2^{(ab)}(t,t') &= C_{ab}(t,t') \\
Q_3^{(ab)}(t,t') &= R_{ab}(t',t) \\
Q_4^{(ab)}(t,t') &= R_{ab}(t,t') \\
Q_5^{(ab)}(t) &= C_{ab}(t,0) \\
Q_6^{(ab)}(t) &= R_{ab}(t,0)
\end{cases}
\end{equation}
Note that $Q_3^{(ab)}(t,t') = Q_4^{(ba)}(t',t)$. Moreover, we claim that, as in the standard case~\cite{sompolinsky1982relaxational}, $Q_1^{(ab)}=0$. The saddle point equations are obtained by differentiating the argument of the exponentials in Eq.~\eqref{eq:Zavg1} with respect to the set of dynamical overlaps $Q_i^{(ab)}$ and they read:
\begin{equation}\label{eq:spparameters}
    \begin{cases}
    iP_1^{(11)} &= \frac{p}{4}Q_2^{(11) p-1} - \gamma\frac{p}{4} \left(Q_2^{(12)p-1} + Q_2^{(21)p-1} \right) + \gamma^2\frac{p}{4} Q_2^{(22)p-1} \\
    iP_2^{(11)} &= 0 \\
    iP_3^{(11)} &= \frac{p}{4}(p-1)Q_4^{(11)}Q_2^{(11)p-2} + \gamma \frac{p}{4}(p-1)Q_4^{(12)}Q_2^{(11)p-2} - \gamma \frac{p}{4}(p-1)Q_4^{(21)}Q_2^{(21)p-2} - \gamma^2 \frac{p}{4}(p-1)Q_4^{(11)}Q_2^{(21)p-2}\\ 
    iP_4^{(11)} &= \frac{p}{4}(p-1)Q_3^{(11)}Q_2^{(11)p-2} + \gamma \frac{p}{4}(p-1)Q_3^{(21)}Q_2^{(11)p-2} - \gamma \frac{p}{4}(p-1)Q_3^{(12)}Q_2^{(12)p-2} - \gamma^2 \frac{p}{4}(p-1)Q_3^{(11)}Q_2^{(12)p-2} \\
    iP_5^{(11)} &= \frac{p}{2T}Q_5^{(11)} - \gamma \frac{p}{2T}Q_5^{(21)} \\
    iP_6^{(11)} &= 0 \\
    \\
    iP_1^{(12)} &= \frac{p}{4}Q_2^{(12)p-1} - \gamma\frac{p}{4} \left(Q_2^{(22)p-1} - Q_2^{(11)p-1} \right) - \gamma^2\frac{p}{4} Q_2^{(22)p-1} \\
    iP_2^{(12)} &= 0 \\
    iP_3^{(12)} &= \frac{p}{4}(p-1)Q_4^{(12)}Q_2^{(12)p-2} - \gamma \frac{p}{4}(p-1)Q_4^{(11)}Q_2^{(12)p-2} - \gamma \frac{p}{4}(p-1)Q_4^{(22)}Q_2^{(22)p-2} + \gamma^2 \frac{p}{4}(p-1)Q_4^{(21)}Q_2^{(22)p-2}\\ 
    iP_4^{(12)} &= \frac{p}{4}(p-1)Q_3^{(12)}Q_2^{(12)p-2} + \gamma \frac{p}{4}(p-1)Q_3^{(11)}Q_2^{(11)p-2} + \gamma \frac{p}{4}(p-1)Q_3^{(22)}Q_2^{(12)p-2} + \gamma^2 \frac{p}{4}(p-1)Q_3^{(21)}Q_2^{(11)p-2} \\
    iP_5^{(12)} &= \frac{p}{2T}Q_5^{(12)} - \gamma \frac{p}{2T}Q_5^{(22)} \\
    iP_6^{(12)} &= 0 \\
    \\
    iP_1^{(21)} &= \frac{p}{4}Q_2^{(21)p-1} - \gamma\frac{p}{4} \left(Q_2^{(11)p-1} - Q_2^{(22)p-1} \right) - \gamma^2\frac{p}{4} Q_2^{(12)p-1} \\
    iP_2^{(21)} &= 0 \\
    iP_3^{(21)} &= \frac{p}{4}(p-1)Q_4^{(21)}Q_2^{(21)p-2} + \gamma \frac{p}{4}(p-1)Q_4^{(11)}Q_2^{(11)p-2} + \gamma \frac{p}{4}(p-1)Q_4^{(22)}Q_2^{(21)p-2} + \gamma^2 \frac{p}{4}(p-1)Q_4^{(12)}Q_2^{(11)p-2}\\ 
    iP_4^{(21)} &= \frac{p}{4}(p-1)Q_3^{(21)}Q_2^{(21)p-2} - \gamma \frac{p}{4}(p-1)Q_3^{(11)}Q_2^{(21)p-2} - \gamma \frac{p}{4}(p-1)Q_3^{(22)}Q_2^{(12)p-2} + \gamma^2 \frac{p}{4}(p-1)Q_3^{(12)}Q_2^{(22)p-2} \\
    iP_5^{(21)} &= \frac{p}{2T}Q_5^{(21)} + \gamma \frac{p}{2T}Q_5^{(11)} \\
    iP_6^{(21)} &= 0 \\
    \\
     iP_1^{(22)} &= \frac{p}{4}Q_2^{(22)p-1} + \gamma\frac{p}{4} \left(Q_2^{(12)p-1} + Q_2^{(21)p-1} \right) + \gamma^2\frac{p}{4} Q_2^{(11)p-1} \\
    iP_2^{(22)} &= 0 \\
    iP_3^{(22)} &= \frac{p}{4}(p-1)Q_4^{(22)}Q_2^{(22)p-2} - \gamma \frac{p}{4}(p-1)Q_4^{(21)}Q_2^{(22)p-2} + \gamma \frac{p}{4}(p-1)Q_4^{(12)}Q_2^{(12)p-2} - \gamma^2 \frac{p}{4}(p-1)Q_4^{(22)}Q_2^{(12)p-2}\\ 
    iP_4^{(22)} &= \frac{p}{4}(p-1)Q_3^{(22)}Q_2^{(22)p-2} - \gamma \frac{p}{4}(p-1)Q_3^{(12)}Q_2^{(22)p-2} + \gamma \frac{p}{4}(p-1)Q_3^{(21)}Q_2^{(21)p-2} - \gamma^2 \frac{p}{4}(p-1)Q_3^{(22)}Q_2^{(21)p-2} \\
    iP_5^{(22)} &= \frac{p}{2T}Q_5^{(22)} + \gamma \frac{p}{2T}Q_5^{(12)} \\
    iP_6^{(22)} &= 0 \\
    \end{cases}
\end{equation}
Note that $iP_2 = iP_6 = 0$ because $Q_1 = 0$ and because of causality, which implies that $Q_3^{(ab)}(t,t')Q_4^{(ab)}(t,t') = 0$. Substitution of the saddle point expressions of Eq. \eqref{eq:spparameters} in Eq. \eqref{eq:Zavgdelta}, together with Eq. \eqref{eq:intdynpar} yields the effective dynamical equations for the spins of the system. Using a vector notation, $\bm{\sigma}(t)=\begin{bmatrix} \sigma^{(1)} & \sigma^{(2)} \end{bmatrix}^T$, we obtain
\begin{equation}\label{eq:effdynpspin}
    \partial_t \bm{\sigma}(t) = -\left(\mathds{1} + \bm{\Gamma}\right) \cdot \bm{\mu}(t) \cdot \bm{\sigma}(t) + \int_0^t d\tau \mathbf{M_R}(t,\tau) \cdot \bm{\sigma}(\tau) + \frac{1}{T}\left(\mathds{1} + \bm{\Gamma} \right) \cdot \mathbf{C}_0(p,t) \cdot \bm{\sigma}(0) + \bm{\xi}(t)
\end{equation}
where $\mu_{ab}(t) \equiv \delta_{ab}\mu^{(a)}(t)$,  $\bm{\Gamma} \equiv \begin{bmatrix}
                            0 & -\gamma \\
                            \gamma & 0
                        \end{bmatrix}$. The correlations of the noise $\bm{\xi}(t)$ are
\begin{equation}
    \left\langle \bm{\xi}(t)\bm{\xi}(t')\right\rangle = 2T\delta(t-t')\mathds{1}_2 + \mathbf{D}(t,t')
\end{equation}
while the expression for the memory kernel $\mathbf{M_R}(t,t')$, the noise correlations $\mathbf{D}(t,t')$ and the coupling with the initial conditions $\mathbf{C}_{0}(t,0)$ 
are the following:
\begin{align}
    \mathbf{M_R}(t,\tau) &= \frac{p(p-1)}{2} \begin{bmatrix}
        C_{11}^{p-2} \left(R_{11} + \gamma R_{12} \right) - \gamma C_{21}^{p-2}\left(R_{21} + \gamma R_{22} \right) &&& C_{12}^{p-2} \left(R_{12} - \gamma R_{11} \right) - \gamma C_{22}^{p-2}\left(R_{22} - \gamma R_{21} \right) \\ \\ 
        C_{21}^{p-2} \left(R_{21} + \gamma R_{22} \right) + \gamma C_{11}^{p-2}\left(R_{11} + \gamma R_{12} \right) &&&
        C_{22}^{p-2} \left(R_{22} - \gamma R_{21} \right) + \gamma C_{12}^{p-2}\left(R_{12} - \gamma R_{11} \right)
    \end{bmatrix} \\
    \mathbf{D}(t,t') &\equiv \left( \mathds{1} + \bm{\Gamma} \right) \cdot \mathbf{C}_0(p,t) \cdot \left( \mathds{1} + \bm{\Gamma^T} \right)  \\ 
    C_{0}(p,t,0)_{ij} &= \frac{p}{2} C_{ij}^{p-1}(t,0) 
\end{align}
From Eq. \eqref{eq:effdynpspin} by using the definition \eqref{eq:corr} and the equalities 
\begin{align}
    R_{ab}(t,t') &= \left\langle \frac{\delta \sigma^{(a)}(t)}{\delta \xi^{(b)}(t')} \right\rangle \\
    \langle \xi^{(a)}(t)\sigma^{(b)}(t')\rangle &= 2T R_{ab}(t',t) + \int d\tau \left[\mathbf{D}(t,\tau) \cdot \mathbf{R^T}(t',\tau)\right]_{ab}
\end{align}
we obtain 
\begin{align}\label{eq:pspin_C_TD}
        \partial_t\mathbf{R}(t,t') &= - \left(\mathds{1} + \bm{\Gamma}\right) \cdot \bm{\mu}(t) \cdot \mathbf{R}(t,t') + \int_{t'}^t d\tau \mathbf{M_R}(t,\tau) \cdot \mathbf{R}(\tau,t') + \delta(t-t')\mathds{1}_2 \\
        \partial_t \mathbf{C}(t,t') &= - \left(\mathds{1} + \bm{\Gamma}\right) \cdot \bm{\mu}(t) \cdot \mathbf{C}(t,t') + \int_0^t d\tau \mathbf{M_R}(t,\tau) \cdot \mathbf{C}(\tau,t') + 2T\mathbf{R}(t',t)\\ &+ \int_0^{t'} d\tau \mathbf{D}(t,\tau) \cdot \mathbf{R^T}(t',\tau)  + \frac{1}{T}\left(\mathds{1} + \bm{\Gamma} \right) \cdot \mathbf{C}_0(p,t) \cdot \mathbf{C}(0,t') \nonumber
\end{align}
By imposing the spherical constraint  $\sum_{i=1}^N \left\langle \left(\sigma_i^{(a)}\right)^2 (t)\right\rangle = N$ for $a=1,2$ one can find a self-consistent expression for the restoring forces $\bm{\mu}(t)$. If we denote the diagonal part of a matrix by $\text{Diag}(\mathbf{A})_{ab} \equiv \delta_{ab}A_{ab}$ the spherical constraint condition reads $\text{Diag}(\mathbf{C}(t,t)) = 1$. By differentiating with respect to $t$ we get 
\begin{equation}
\lim_{t'\rightarrow t}\text{Diag}\left(\partial_t\mathbf{C}(t,t') + \partial_{t'}\mathbf{C}(t,t')  \right) = \lim_{t'\rightarrow t}\text{Diag}\left(\partial_t\mathbf{C}(t,t') + \partial_{t'}\mathbf{C^T}(t',t)\right) =  0 
\end{equation}
which leads to 
\begin{align}
\begin{split}
    \bm{\mu}(t) &= \text{Diag}\Biggl[T\mathds{1}_2 + \int_0^t \dd \tau \mathbf{M}(t,\tau) \cdot     \mathbf{C^T}(t,\tau) + \int_{0}^{t} d\tau  \mathbf{D}(t,\tau) \cdot \mathbf{R^T}(t,\tau) \\ &+ \frac{1}{T}\left(\mathds{1} + \bm{\Gamma} \right) \cdot \mathbf{C}_0(p,t) \cdot \mathbf{C^T}(t,0)\Biggr]
\end{split}
\end{align}
Since the dynamics of the system starts within the steady state, we assume that time translation invariance holds: $\mathbf{C}(t,t') \equiv \mathbf{C}(t-t')$ and $\mathbf{R}(t,t') \equiv \mathbf{R}(t-t')$ The equations of motion for the response and correlation matrices therefore read, when taking $t'=0$
\begin{align}
        \partial_t\mathbf{R}(t) &= - \left(\mathds{1} + \bm{\Gamma}\right) \cdot \bm{\mu}(t) \cdot \mathbf{R}(t) + \int_{t'}^t \dd\tau \mathbf{M_R}(t-\tau) \cdot \mathbf{R}(\tau) + \delta(t)\mathds{1}_2 \label{eq:pspin_R_TD_TTI}\\
        \partial_t \mathbf{C}(t) &= -\left(\mathds{1} + \bm{\Gamma}\right) \cdot \bm{\mu}(t) \cdot \mathbf{C}(t) + \int_0^t \dd \tau \mathbf{M_R}(t-\tau) \cdot \mathbf{C}(\tau) + \frac{1}{T}\left(\mathds{1} + \bm{\Gamma} \right) \cdot \mathbf{C}_0(p,t)  \label{eq:pspin_C_TD_TTI} \\
        \begin{split}\label{eq:pspin-mu-TD-TTI}
        \bm{\mu}(t) &= \text{Diag}\Biggl[T\mathds{1}_2 + \int_0^t \dd\tau \mathbf{M}(t - \tau) \cdot \mathbf{C^T}(t - \tau) + \int_{0}^{t} \dd\tau \mathbf{D}(t - \tau) \cdot \mathbf{R^T}(t - \tau)  \\
        &+ \frac{1}{T}\left(\mathds{1} + \bm{\Gamma} \right) \cdot \mathbf{C}_0(p,t) \cdot \mathbf{C^T}(t)\Biggr] 
        \end{split}
\end{align}
For numerical purposes it is more convenient to work with the integrated response matrix $\mathbf{F}(t,t')$ \cite{kim2001dynamics}, defined by
\begin{equation}
    F_{ab}(t,t') \equiv -\int_{t'}^{t}d\tau R_{ab}(t,\tau)
\end{equation}
since it has a smoother relaxation than $\mathbf{R}(t,t')$. The equations of motion in the time translation invariant regime are
\begin{align}
        \partial_t\mathbf{F}(t) &= - \mathds{1} - \left(\mathds{1} + \bm{\Gamma}\right) \cdot \bm{\mu}(t) \cdot \mathbf{F}(t) + \int_{t'}^t d\tau \mathbf{M_F}(t-\tau) \cdot \mathbf{F}(\tau) \label{eq:pspin_F_TD_TTI}\\
        \partial_t \mathbf{C}(t) &= -\left(\mathds{1} + \bm{\Gamma}\right) \cdot \bm{\mu}(t) \cdot \mathbf{C}(t) + \int_0^t d\tau \mathbf{M_F}(t-\tau) \cdot \mathbf{C}(\tau) + \frac{1}{T}\left(\mathds{1} + \bm{\Gamma} \right) \cdot \mathbf{C}_0(p,t)  \label{eq:pspin_CF_TD_TTI} \\
        \begin{split}\label{eq:pspin-muF-TD-TTI}
        \bm{\mu}(t) &= \text{Diag}\Biggr[T\mathds{1}_2 + \int_0^t d\tau \mathbf{M_F}(t - \tau) \cdot \mathbf{C^T}(t - \tau) + \int_{0}^{t} d\tau \mathbf{D}(t - \tau) \cdot \partial_\tau\mathbf{F^T}(t - \tau)  \\
        &+  \frac{1}{T}\left(\mathds{1} + \bm{\Gamma} \right) \cdot \mathbf{C}_0(p,t) \cdot \mathbf{C^T}(t)\Biggr] 
        \end{split}
\end{align}
and the memory kernel $\mathbf{M_F}(t)$ that reads
\begin{equation}
    \begin{split}
        \mathbf{M}_\mathbf{F}(t-\tau) &\equiv -\frac{p(p-1)}{2} \times \\&\begin{bmatrix} 
            C_{11}^{p-2} \left(\partial_\tau F_{11} + \gamma \partial_\tau F_{12} \right) - \gamma C_{21}^{p-2}\left(\partial_\tau F_{21} + \gamma \partial_\tau F_{22} \right) && C_{12}^{p-2} \left(\partial_\tau F_{12} - \gamma \partial_\tau F_{11} \right) - \gamma C_{22}^{p-2}\left(\p_\tau F_{22} - \gamma \p_\tau F_{21} \right) \\ \\ 
            C_{21}^{p-2} \left(\p_\tau F_{21} + \gamma \p_\tau F_{22} \right) + \gamma C_{11}^{p-2}\left(\p_\tau F_{11} + \gamma \p_\tau F_{12} \right) && 
            C_{22}^{p-2} \left(\p_\tau F_{22} - \gamma \p_\tau F_{21} \right) + \gamma C_{12}^{p-2}\left(\p_\tau F_{12} - \gamma \p_\tau F_{11} \right)
        \end{bmatrix}
    \end{split}
\end{equation}
the companion initial conditions are $\mathbf{C}(0) \equiv \mathds{1}$, $\mathbf{F}(0) = \mathbf{0}$. Together with Eq.~\eqref{eq:effdynpspin}, these equations correspond to~\eqrefs{eq:effestocheq, eq:F, eq:C, eq:pspin-mu_F-TD-TTI} presented in the main text.

In the following section we provide the reader with some details on the algorithm implemented to integrate the dynamical equations.

\section{Numerical Integration of the dynamical equations}\label{app:numerics}

We present here the integration scheme adopted to numerically solve the dynamical equations of motion, \eqrefs{eq:pspin_F_TD_TTI, eq:pspin_CF_TD_TTI, eq:pspin-muF-TD-TTI}. The basic idea for the algorithm was first given in \cite{fuchs1991comments} in the context of the mode coupling equations. It was then applied to the aging of the $p-$spin model in \cite{kim2001dynamics} and generalised in \cite{berthier2007spontaneous} for the case of non stationary systems. A pedagogical description of the algorithm can be found in \cite{folena2020mixed}.

To integrate \eqrefs{eq:pspin_F_TD_TTI, eq:pspin_CF_TD_TTI, eq:pspin-muF-TD-TTI}, we take a grid of $M$ points with a discretisation step $\Delta t$, so that any function $f(t)$ can be discretised as $f_k \equiv f(t_k)$ with $t_k \equiv k\Delta t$. Suppose that we know the first $n$ value of the correlation and the integrated response along the discretized line and that we want to know their value at the step $n+1$. By resorting to finite difference approximations, we can obtain a set of non-linear equation for the variable $\mathbf{V}_i \equiv \begin{bmatrix} \mathbf{C}_i & \mathbf{F}_i \end{bmatrix}$ of the form
\begin{equation}
    \mathbf{V}_{n+1} = \mathbf{NL}\left(\mathbf{V}_0,\mathbf{V}_1,...,\mathbf{V}_n, \mathbf{V}_{n+1}\right) 
\end{equation}
where $\mathbf{NL}$ is a generic nonlinear operator. This set of equation can be solved iteratively starting from the initial condition $\mathbf{V}_{n+1} = \mathbf{V}_{n}$. Once convergence is achieved, one can move to the next point along the discretised line. This is the propagation step.

Once all the values of $\mathbf{C}$ and $\mathbf{F}$ have been obtained along the grid, one performs a decimation and a rescaling to compress the information obtained and reach longer times. Half of the points are discarded by performing the substitution $\mathbf{V}_{2i} \rightarrow \mathbf{V}_i$ for $i=0,...,M-1$ and the time step is rescaled by a factor two, $\Delta t \rightarrow 2\Delta t$. The propagation step is then repeated starting from the point $n+1 = \frac{M}{2}$. Large times can be quickly achieved by iterating these two steps.

The finite difference approximations that we adopt are the following. For the left hand side of~\eqrefs{eq:pspin_F_TD_TTI,eq:pspin_CF_TD_TTI} the time derivative of a function at step $k+1$, $\partial_t f_{k+1}$ can be approximated as 
\begin{equation}
    \p_t f_{k+1} = \frac{1}{2\Delta t}\left(3f_{k+1} - 4f_{k} + f_{k-1} \right) + O(\Delta t^2)
\end{equation}
while the integrals on the right hand side have the general form 
\begin{align}
    I_{n+1} &\equiv \int_0^{t_{n+1}}\dd\tau A(t-\tau)\partial_\tau B(t-\tau) C(\tau) \\
    J_{n+1} &\equiv \int_0^{t_{n+1}} \dd\tau A(t-\tau)\partial_\tau B(t-\tau) C(\tau)
\end{align}
and they can be approximated as
\begin{align}
    I_{n+1} &= \sum_{j=0}^{n} \frac{1}{4} \left(A_{n+1-j} + A_{n-j}\right)\left(B_{n-j} - B_{n+1-j}\right)\left(C_j + C_{j+1}\right) + O(\Delta t^2) \\
    J_{n+1} &= \sum_{j=0}^{n} \frac{1}{4} \left(A_{n+1-j} + A_{n-j}\right)\left(B_{n-j} - B_{n+1-j}\right)\left(C_{n+1-j} + C_{n-j}\right) + O(\Delta t^2) \\
\end{align}
For the results presented in the main text we have adopted a grid of $M= 1024$ points and an initial timestep of $\Delta t = 10^{-10}$.  

\section{aFDT for the IO dynamics of the $p-$spin}\label{app:FDTHO}

Based on the results for the correlation and the response of an harmonic oscillator we make the following $\textit{ansatz}$ concerning the form of the violation of the FDT for the Twisted Dynamics of the $p-$spin:
\begin{equation}\label{eq:FDT_gamma_pspin}
    \frac{1}{T}\left[\mathbf{C}(t) - \mathbf{C}(0) \right] = \mathbf{F}(t) \cdot \left( \mathds{1} + \bm{\Gamma} \right)
\end{equation}
that is, an analogous form of the FDT relation holds if one looks at the proper combination of response functions. We call this the aFDT relation. In this section we will show that this relation is indeed satisfied by~\eqrefs{eq:pspin_F_TD_TTI, eq:pspin_CF_TD_TTI, eq:pspin-muF-TD-TTI}. The strategy of the proof is the following: starting from \eqrefs{eq:pspin_CF_TD_TTI, eq:pspin-muF-TD-TTI} and the aFDT we will obtain Eq.~\eqref{eq:pspin_F_TD_TTI}. To do so, we first rewrite Eq.~\eqref{eq:FDT_gamma_pspin} in a differential form:
\begin{equation}\label{eq:FDT_gamma_pspin_differential}
    \frac{1}{T}\partial_t\mathbf{C}(t) = \partial_t\mathbf{F}(t) \cdot \left(\mathds{1} + \bm{\Gamma} \right)
\end{equation}
using Eq.~\eqref{eq:FDT_gamma_pspin_differential} we can rewrite the memory kernel $\mathbf{M_F}$ in a simpler form: 
\begin{equation}\label{eq:MF_FDTgamma_pspin}
    \mathbf{M_F}(t-\tau) = \frac{1}{T}\left( \mathds{1} + \bm{\Gamma} \right) \cdot \partial_\tau \mathbf{C}_0(t-\tau)
\end{equation}
We now proceed to compute the left hand side of Eq.~\eqref{eq:FDT_gamma_pspin_differential} using Eq.~\eqref{eq:pspin_C_TD_TTI}:
\begin{align}
\begin{split} \label{eq:pspin_C_aFDT}
    \frac{1}{T}\mathbf{C}(t) &= -\frac{1}{T}\left(\mathds{1} + \bm{\Gamma} \right) \cdot \bm{\mu}(t) \cdot \mathbf{C}(t) + \frac{1}{T}\int_0^t d\tau \mathbf{M_F}(t-\tau)\cdot\mathbf{C}(\tau) + \frac{1}{T^2}\left(\mathds{1} + \bm{\Gamma} \right) \cdot \mathbf{C}_0(t) \\
    &= -\left(\mathds{1} + \bm{\Gamma}\right) \cdot \left(\frac{1}{T}\mathds{1} + \mathbf{F}(t)\cdot\left(\mathds{1} + \bm{\Gamma}\right) \right) + \frac{1}{T}\int_0^t d\tau \mathbf{M_F}(t-\tau)\cdot\mathbf{C}(\tau) + \frac{1}{T^2}\left(\mathds{1} + \bm{\Gamma} \right) \cdot \mathbf{C}_0(t) 
    \end{split} \\
    \begin{split}
    \bm{\mu}(t) &= \text{Diag} \Biggl[T\mathds{1}_2 + \int_0^t d\tau \mathbf{M_F}(t - \tau) \cdot \mathbf{C^T}(t - \tau) + \int_{0}^{t} d\tau \mathbf{D}(t - \tau) \cdot \partial_\tau\mathbf{F^T}(t - \tau)  \\
    &+ \frac{1}{T}\left(\mathds{1} + \bm{\Gamma} \right) \cdot \mathbf{C}_0(p-1,t) \cdot \mathbf{C^T}(t)\Biggr]
    \end{split}
\end{align}
to evaluate the time integral, we can use Eq. \eqref{eq:MF_FDTgamma_pspin}, integration by parts, the aFDT relation and the boundary condition $\mathbf{C}(0) = \mathds{1}$, $\mathbf{F}(0) = \mathbf{0}$ to obtain
\begin{align}
    \frac{1}{T}\int_0^t d\tau \mathbf{M_F}(t-\tau)\cdot \mathbf{C}(\tau) &= \frac{1}{T^2}\left(\frac{p}{2}\mathds{1} - \mathbf{C}_0(t) \right) + \int_0^t d\tau \mathbf{M_F}(t - \tau) \cdot \mathbf{F}(\tau) \cdot \left( \mathds{1} + \bm{\Gamma} \right) \\
    \int_{0}^{t} d\tau \mathbf{D}(t - \tau) \cdot \partial_\tau\mathbf{F^T}(t - \tau) &= \frac{1}{2T}\left(\mathds{1} + \bm{\Gamma}\right) \cdot \int_0^t d\tau \mathbf{C}_0 \cdot \p_\tau \mathbf{C^T}(\tau) \\
    \int_0^t d\tau \mathbf{M_F}(t - \tau) \cdot \mathbf{C^T}(t - \tau) &= \frac{1}{T}\left[ \frac{p}{2}\mathds{1} - \mathbf{C}_0(t) \cdot \mathbf{C^T}(t) - \frac{1}{T}\int_0^{t} d\tau \mathbf{C}_0(t-\tau)\cdot \p_\tau \mathbf{C^T}(\tau)\right] 
\end{align}
So that $\bm{\mu}$ reads after simplifying and taking the diagonal part:
\begin{equation}
    \bm{\mu}(t) = \left(T + \frac{p}{2T}\right)\mathds{1} \equiv \bm{\mu}_{\infty}
\end{equation}
putting everything back in Eq.~\eqref{eq:pspin_C_aFDT} and simplifying we get that
\begin{equation}
    \frac{1}{T}\partial_t \mathbf{C}(t) = \p_t\mathbf{F}(t) \cdot \left( \mathds{1} + \bm{\Gamma}\right) 
\end{equation}
with $\p_t\mathbf{F}(t)$ given by Eq.~\eqref{eq:pspin_F_TD_TTI}, and the proof is completed.
\twocolumngrid

\providecommand{\noopsort}[1]{}\providecommand{\singleletter}[1]{#1}%

\end{document}